\documentclass[preprint,12pt, review]{elsarticle}
\usepackage{amssymb}
\usepackage{graphicx}
\usepackage{tabularx}
\usepackage{amsthm}
\usepackage{amsmath}
\usepackage{framed} % Framing content
\usepackage{etoolbox}
\usepackage{multicol} % Multiple columns environment
\usepackage{setspace}
\usepackage{nomencl} % Nomenclature package
\usepackage{eqnarray}
\usepackage{xcolor}
\usepackage{float}
\usepackage[section]{placeins}
\usepackage{subcaption}
\usepackage[colorlinks=true,bookmarks=false,citecolor=blue,urlcolor=blue]{hyperref}
\def\equationautorefname~#1\null{%
  Equation~(#1)\null
}
\usepackage{xspace}

\newcommand\review[1]{\textcolor{black}{#1}}
\newcommand\voa{{{VOA\xspace}}}
\newcommand\ica{{{ICA\xspace}}}
\newcommand\fua{{{FUA\xspace}}}
\newcommand\massflow{{$\dot{m}$}}
\newcommand\twodim{two-dimensional}
\newcommand\threedim{three-dimensional}
\usepackage{lineno}
\usepackage{booktabs}

\journal{Computers and Fluids}

\begin{document}

\begin{frontmatter}

%% Title, authors and addresses

%% use the tnoteref command within \title for footnotes;
%% use the tnotetext command for theassociated footnote;
%% use the fnref command within \author or \address for footnotes;
%% use the fntext command for theassociated footnote;
%% use the corref command within \author for corresponding author footnotes;
%% use the cortext command for theassociated footnote;
%% use the ead command for the email address,
%% and the form \ead[url] for the home page:
%% \title{Title\tnoteref{label1}}
%% \tnotetext[label1]{}
%% \author{Name\corref{cor1}\fnref{label2}}
%% \ead{email address}
%% \ead[url]{home page}
%% \fntext[label2]{}
%% \cortext[cor1]{}
%% \affiliation{organization={},
%%             addressline={},
%%             city={},
%%             postcode={},
%%             state={},
%%             country={}}
%% \fntext[label3]{}

% \title{Topology optimization for manifold design based on flow uniformity}
% \title{Comparative assessment of density-based topology optimization strategies for optimal design of flow manifolds using SU2.}
\title{Density-based topology optimization strategy for optimal design of uniform flow manifolds.}

% Density-based fluid topology optimization using adjoint-based in-compressible flow solver in SU2.

%% use optional labels to link authors explicitly to addresses:
%% \author[label1,label2]{}
%% \affiliation[label1]{organization={},
%%             addressline={},
%%             city={},
%%             postcode={},
%%             state={},
%%             country={}}
%%
%% \affiliation[label2]{organization={},
%%             addressline={},
%%             city={},
%%             postcode={},
%%             state={},
%%             country={}}

\author[inst1,inst2]{Sanjay Vermani}

\affiliation[inst1]{organization={Energy Systems and Components Optimization, Water \& Energy Transition Unit}, Flemish Institute for Technological Research (VITO),%Department and Organization
            addressline={Boeretang~200}, 
            city={Mol},
            postcode={2400}, 
            country={Belgium}}

\author[inst1,inst2,inst3]{Nitish Anand}
\ead{nitish.anand@vito.be}
\affiliation[inst2]{organization={EnergyVille},%Department and Organization
            addressline={Thor Park 8310}, 
            city={Genk},
            postcode={3600}, 
            country={Belgium}}
\affiliation[inst3]{organization={Delft University of Technology},%Department and Organization
            addressline={Kluyverweg 1}, 
            city={Delft},
            postcode={2629HS}, 
            country={The Netherlands}}

% ABSTRACT

\begin{abstract}

% Uniform flow distribution across parallel channels directly impacts the performance and efficiency of many fluid and energy systems. 

Flow manifolds are devices that distribute or collect fluid across multiple channels, playing a crucial role in the performance of many fluid and energy systems. However, designing efficient manifolds that ensure uniform flow distribution remains challenging, especially for multi-channel three-dimensional manifolds. This study presents a scalable topology optimization framework for systematically designing multi-channel flow manifolds. The proposed method builds on the conventional density-based topology optimization formulation by introducing a flow maldistribution coefficient as an explicit constraint. This novel approach was implemented using the incompressible Navier-Stokes flow solver available in the open-source CFD suite SU2. The performance of the proposed method was benchmarked against two established topology optimization strategies using an exemplary planar z-type flow manifold, where both the inlet and outlet manifolds were designed simultaneously. The results demonstrate that the proposed method achieves flow uniformity comparable to established approaches while significantly reducing computational costs. Furthermore, when applied to large-scale three-dimensional problems, the proposed method produces feasible designs that achieve uniform flow distribution and exhibit innovative geometrical features. These results highlight the robustness and scalability of the proposed method.

% Thus confirming its effectiveness, despite relying on a single global constraint. To assess performance in three-dimensional domains, the method was applied to two cases: parallel Z-flow manifolds and vertical radial manifolds. The resulting geometries featured internal flow paths and asymmetric 3D shapes that met strict flow uniformity requirements without outlet-specific constraints. These results demonstrate that the proposed approach maintains flow distribution performance while reducing constraint complexity, enabling efficient design optimization of flow manifolds in high-dimensional and large-scale systems.
\end{abstract}

%%Graphical abstract
\begin{graphicalabstract}
\vspace{1cm}
\includegraphics[width=1.0\textwidth]{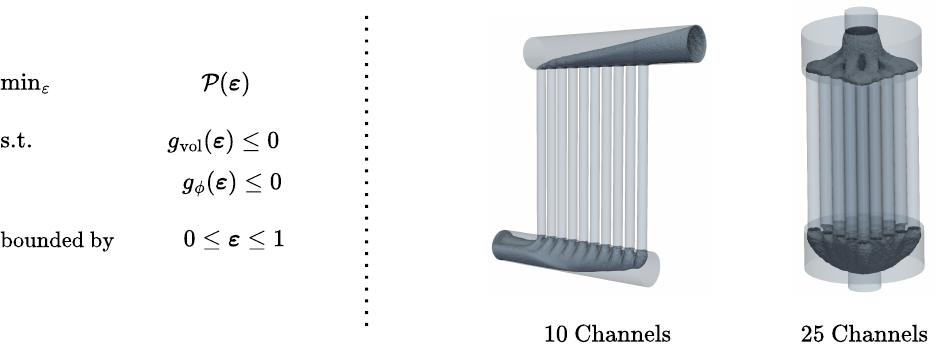}
\end{graphicalabstract}

%%Research highlights
\begin{highlights}
\item An adjoint-based fluid topology optimization framework using open-source simulation suite \textit{SU2}.
\item A novel density-based topology optimization formulation for multi-channel flow manifold design.
\item Application to complex \threedim\  multi-channel flow manifold design cases.
\end{highlights}

\begin{keyword}
%% keywords here, in the form: keyword \sep keyword
Topology optimization \sep Flow manifolds \sep Uniform flow distribution \sep Adjoint-based optimization \sep SU2
%% PACS codes here, in the form: \PACS code \sep code
% \PACS 0000 \sep 1111
%% MSC codes here, in the form: \MSC code \sep code
%% or \MSC[2008] code \sep code (2000 is the default)
% \MSC 0000 \sep 1111
\end{keyword}

\end{frontmatter}

\makenomenclature

\renewcommand{\nomgroup}[1]{%
  \ifthenelse{\equal{#1}{A}}{\item[\textbf{General Symbols}]}{}%
  \ifthenelse{\equal{#1}{B}}{\item[\textbf{Greek Symbols}]}{}%
  \ifthenelse{\equal{#1}{C}}{\item[\textbf{Subscripts}]}{}%
  \ifthenelse{\equal{#1}{D}}{\item[\textbf{Superscripts}]}{}%
}

\setlength{\columnsep}{1cm}

\setlength{\nomitemsep}{-\parsep} % Baseline skip between items

\renewcommand*\nompreamble{\begin{multicols}{3}}

\renewcommand*\nompostamble{\end{multicols}}

% \input{Tex/7_Nomenclature}

%\linenumbers

%% main text
% INTRODUCTION
% \clearpage
\section{Introduction}

Manifolds are critical components in fluid-flow systems, enabling controlled and efficient distribution of fluid across different flow paths. Typically, they fall into two categories: combining manifolds, which merge multiple incoming flows into a single outlet, and distributing manifolds (also called flow distributors), which channel fluid from one or more inlets into multiple flow paths~\cite{WANG20111331}. The primary goal of manifolds is to ensure uniform or design-specific flow distribution across the channels. They play a crucial role in various applications, such as in micro-channel heat sinks~\cite{maldist_heatsink}, battery cold plates~\cite{bcoldplate}, reactors~\cite{maldistrib_reactor}, fuel cells~\cite{maldistrib_fuelcell}, heat exchangers~\cite{LALOT1999847, maldistrib_hex}, and gas treatment units~\cite{maldistriub_chemical}, among others. An inefficient manifold design can result in poor heat transfer, temperature control issues, and increased pumping power~\cite{maldistrib_hex, SIDDIQUI2017969}. Therefore, optimizing the design of manifolds is essential to improve the overall efficiency of fluid-flow systems.

Manifold design has traditionally relied on empirical approaches, most notably London’s 1968 method~\cite{London}, which is based on Bernoulli’s principle under the assumption of one-dimensional, uniform flow. This method yields a simple one-dimensional design formula that links the inlet manifold’s curvature to the inlet and outlet lengths, assuming a fixed box-type outlet manifold. Since its inception, the characteristic tapered shape for the inlet manifold, obtained using London’s method, has been widely adopted for applications with parallel channels~\cite{KIM1995329, dhariya, TONG20093552}. These studies advocate the effectiveness of London’s design in promoting uniform flow distribution.

Despite its simplicity and popularity, London’s approach exhibits several limitations. It assumes one-dimensional velocity and pressure distributions, a fixed outlet manifold geometry, and neglects key effects such as momentum and viscous losses, which are critical for accurate performance prediction. To address these shortcomings, several researchers have extended London’s model. For instance, Refs.~\cite{bajura} and~\cite{yang} incorporated axial momentum loss in oblique dividing manifolds using control volume techniques. These methods yield parametric equations that relate geometry and operating conditions to outlet mass flow distribution, offering improved accuracy for simple geometries with constant cross-sections~\cite{yang}.

Beyond analytical extensions, numerous studies have leveraged both experimental and computational fluid dynamics~(CFD) methods to refine flow distribution in manifold systems. For example, Ref.~\cite{jiang} explored the effect of the area ratio and Reynolds number on flow uniformity, demonstrating that a smaller area ratio between the total cross-sectional area of the channels and the manifold’s cross-section can significantly improve flow distribution. Similarly, a review by Ref.~\cite{SIDDIQUI2017969} examined the impact of manifold geometry on flow maldistribution, leading to an analytical model that quantifies the relationship between these two factors. Additionally, Ref.~\cite{GHANI20171143} investigated how different manifold shapes and flow configurations impact flow distribution. The results reported highlight that symmetric triangular manifold geometry, for both combining and dividing manifolds, generates superior flow distribution for z-type flow arrangements. For c-type flow arrangements, a triangular dividing manifold, in conjunction with a trapezoidal combining manifold, yields the best results. These studies typically vary only a limited set of geometric parameters under fixed assumptions, failing to explore the full range of the available manifold design space.

As manifold systems grow in complexity, classical design theories become inadequate for capturing detailed flow behavior. These methods often fall short in representing intricate flow phenomena such as vortex formation, flow separation, and reattachment, particularly in configurations involving multiple branches or T-junctions~\cite{CFD_need_1, CFD_need_2}. To overcome these limitations, there is an increased interest in CFD-based automated design optimization techniques. 

Generally, CFD-based approaches integrate optimization algorithms with flow solvers to systematically explore designs based on the cost functions and the design variables provided. They not only enhance design accuracy by resolving complex flow physics but also offer the flexibility to handle diverse, intricate geometries. Thanks to advances in numerical methods and the accessibility of high-performance computing, such techniques have been successfully applied in diverse domains, including turbomachinery blades~\cite{Agromayor2022}, aircraft components~\cite{Adler2023b}, wings~\cite{Elham_2018}, and heat exchangers~\cite{PAIRAIKAR2025126230}. CFD-based optimization methods can be broadly categorized into shape optimization~\cite{RobertoJEGTP} and fluid topology optimization (FTO)~\cite{ALLAIRE20211}. While both shape optimization and topology optimization are valuable for manifold design, this paper focuses on the latter, specifically density-based FTO. 

The design of manifolds has benefited from advancements in FTO. For instance, Ref.~\cite{Kubo2016} employed a two-dimensional FTO with level sets to design plate-type manifolds in z-type and u-type configurations. The optimization aimed to minimize pressure drop and improve flow uniformity by using dissipated power as the objective function. Similarly, Ref.~\cite{Zeng2018} applied two-dimensional FTO to design manifolds for both uniform and non-uniform flow distributors, including a constraint on the mass flow threshold in each channel. This method was used to design a two-dimensional inlet manifold for a four-channel radiator, resulting in tapered manifolds similar to those obtained using London’s approach. In Ref.~\cite{MEN2024122008}, a density-based FTO was used to design both inlet-distributing and outlet-combining manifolds for a multi-channel heat sink. Here, the objective was to minimize the temperature in the channels while maintaining an allowable pressure drop. In addition to two-dimensional approaches, some studies have also explored three-dimensional manifold designs using FTO. For example, Ref.~\cite{Dilgen2018} optimized a turbulent \threedim\ manifold with one inlet and three outlets, using dissipated power as the objective while imposing constraints on volume fraction and mass flow in each outlet. Similarly, Ref.~\cite{Roeland} simulated an inlet manifold with three outlets, where the objective was power dissipation, and the mass flow at the outlets was imposed as a boundary condition.

In the majority of these design studies, power dissipation is the primary objective function used in optimization formulations. To ensure flow uniformity, these methods typically impose constraints on the mass flow through each channel. While this approach is intuitive, it becomes increasingly complex as the number of outlets increases since the number of constraints scales linearly with the number of outlets. This leads to a higher number of sensitivity computations required per optimization step, making optimization less efficient for manifolds with a large number of channels. As an alternative, studies by Refs.~\cite{Mayank2023, Manifold_combi_obj_2} investigated the use of a combined objective function that incorporates both power dissipation and flow uniformity. The results showed that the choice of weights for these objective functions significantly influenced the final design, which can thus result in a costly trial-and-error approach. Although these studies highlight significant advancements in manifold design, the development of a scalable FTO strategy that ensures uniform flow distribution across complex manifold systems remains a new research frontier.

Stemming from the above consideration, the objective of this research is to propose a scalable FTO design strategy that enables the realization of manifold geometries with uniform flow distribution. To achieve this, porosity source terms enabling density-based FTO were implemented within the incompressible Navier-Stokes solver available in the open-source CFD tool \textit{SU2}~\cite{su2}. Furthermore, adjoint-based gradient computation of the objective function was enabled by incorporating these source terms into the discrete adjoint solver of \textit{SU2}. A python-based optimization framework, \textit{FlowForge}, was developed to integrate a gradient-based optimization algorithm with both the primal and adjoint solvers. This framework was configured with different commonly used optimization strategies to benchmark the proposed strategy. For this comparative study, the inlet and outlet manifolds of an exemplary planar z-type manifold case, consisting of ten parallel channels, were optimized. Additionally, to assess the scalability of the proposed method to three-dimensional cases, the proposed design strategy was applied to optimize the inlet and outlet manifolds for two commonly used manifold configurations. These configurations include cylindrical z-type parallel flow manifolds~\cite{WANG20111331, bcoldplate} and radial flow manifolds~\cite{radial_case_Ref_1, radial_case_Ref_2}.
 
% \cite{SU2TopOpt} - complain design of airfoils.
% \cite{Elham_2018} - topological optimization of aircraft wing
% \cite{Ole2024} - acoustic design
% \cite{Alonso2023} - design of rotating diffusers

% \cite{Dilgen2018} - 3D flow manifold with turbulence
% \cite{Roeland} - 1 - 3 manifold where the three outlets are on the same plane.
% \cite{Mayank2023}

% METHODOLOGY
\section{Governing equations}
\subsection{Primal solver}
To solve the FTO problem, the fluid flow in the design domain is governed by the incompressible Navier-Stokes equations~\cite{SU2_INC}. These conservation equations can be expressed in the differential form as
\begin{eqnarray}
\label{eq:NS}
\frac{\partial \boldsymbol{V}}{\partial t}+\nabla \cdot \boldsymbol{\mathcal{F}}^c(\boldsymbol{V})-\nabla \cdot \boldsymbol{\mathcal{F}}^v(\boldsymbol{V}, \nabla \boldsymbol{V}) - \boldsymbol{\mathcal{S}} =0,
\end{eqnarray}
where $\boldsymbol{V}$ is the vector of the conservative variables for mass, momentum, and energy equations, and \review{$\boldsymbol{\mathcal{S}}$ represents the source terms vector}. The vector $\boldsymbol{V}$ is given as
\begin{equation}
\boldsymbol{V}=\left\{\rho, \rho \boldsymbol{v}, \rho c_\text{p} T\right\}^{\top}.
% V=\{p, \bar{v}, T\}^{\top}.
\end{equation}
In the above expression, $\rho$ is the fluid density, $\boldsymbol{v}$ is the velocity vector, $T$ is the temperature, and $c_\text{p}$ is the specific heat capacity at constant pressure.

Further, the convective and the viscous fluxes in the equation \eqref{eq:NS} are given as
\begin{align}
\boldsymbol{\mathcal{F}}^c(\boldsymbol{V}) &= 
\left\{
\begin{array}{c}
\rho \boldsymbol{v} \\
\rho \boldsymbol{v} \otimes \boldsymbol{v} + \overline{\boldsymbol{I}} p \\
\rho c_\text{p} T \boldsymbol{v}
\end{array}
\right\},
&
\boldsymbol{\mathcal{F}}^v(\boldsymbol{V}, \nabla \boldsymbol{V}) = 
\left\{
\begin{array}{c}
\cdot \\
\boldsymbol{v}{\overline{\boldsymbol{\tau}}} \\
\kappa \nabla T
\end{array}
\right\},
\end{align}
where $p$ is the static pressure, $\kappa$ is the thermal conductivity of the fluid, and the viscous stress tensor $\overline{\boldsymbol{\tau}}$ is defined as
\begin{equation}
\label{eq:eq4}
\overline{\boldsymbol{\tau}} =\mu\left(\nabla \boldsymbol{v}+\nabla \boldsymbol{v}^{\top}\right)-\mu \frac{2}{3} \overline{\boldsymbol{I}}(\nabla \cdot \boldsymbol{v}),
\end{equation}
where $\mu$ is the dynamic viscosity of the fluid.

To enable topology optimization, the influence of porous material on fluid dynamics is represented by the Brinkman penalization term, as defined in Ref.~\cite{Dilgen2018}. According to this approach, a momentum source term is introduced to account for the resistance caused by solid/porous material. This source term, denoted by $\boldsymbol{\mathcal{S}}$ in equation~\eqref{eq:NS}, takes the form
\begin{equation}
\label{eq:source_term}
\boldsymbol{\mathcal{S}}=\left\{\begin{array}{c}
\cdot \\
\rho \alpha(\varepsilon) \boldsymbol{v}\\
\cdot
\end{array}\right\},
\end{equation}
where $\varepsilon$ is the porosity field within the flow domain. The value of this porosity ranges from $0$ for solid regions to $1$ for fluid regions, with intermediate values representing porous materials that exhibit varying flow resistance. The function $\alpha(\varepsilon)$ represents inverse permeability and is often defined using an interpolation function~\cite{parallel_pipe}. In this study, $\alpha(\varepsilon)$ is defined as
\begin{equation}
\label{eq:simp} % why are we not citing simp?
    \alpha(\varepsilon) = \alpha_{\text{s}} \; q \frac{(1 - \varepsilon)}{q + \varepsilon}.
\end{equation} 
In equation~\eqref{eq:simp}, $q$ is a constant that characterizes the curvature of the function $\alpha(\varepsilon)$, thereby determining the penalization associated with intermediate values of $\varepsilon$. 

Previous research has demonstrated that the convergence of the optimization problem can be sensitive to the selection of the parameter $q$~\cite{Signmond2018}. To address this, the curvature parameter $q$ is gradually increased in steps during the optimization, starting from 0.01 to 0.1, and finally to 1. The update of parameter $q$ is based on a convergence criterion implemented in the optimization problem. Hence, gradually increasing $q$ in this way prevents the algorithm from converging to suboptimal local solutions~\cite{Dilgen2018,parallel_pipe}.

Parameter $\alpha_\text{s}$, in equation \eqref{eq:simp}, represents the maximum penalization for the solid regions, that is, $\varepsilon = 0$. To effectively characterize the impermeable solid regions, this parameter must be sufficiently high; otherwise, it would lead to flow diffusion through the porous media. On the contrary, very high values of $\alpha_\text{s}$ may lead to numerical instability and thus a poor solver convergence~\cite{Dilgen2018}. Therefore, following the recommendation of Ref.~\cite{Signmond2018}, the value of $\alpha_\text{s}$ is determined based on the Darcy number ($\mathrm{Da}$), given as
\begin{equation}
\label{eq:darcy}
\mathrm{Da} = \frac{\mu}{\rho \; \alpha_\text{s} \; a^2},
\end{equation}
which represents the ratio of viscous forces to the Darcy damping forces, with $a$ denoting the characteristic length scale of the problem.

The above-mentioned governing equations, \eqref{eq:NS}-\eqref{eq:eq4}, were already available within the \textit{SU2} suite~\cite{su2}. To enable density-based topology optimization, the \textit{SU2} suite was extended with equation \eqref{eq:source_term}-\eqref{eq:darcy}. The system of partial differential equations is discretized through a finite volume method, where the convective fluxes are reconstructed utilizing the Flux-Difference Splitting (FDS) technique~\cite{Toro1997}. To attain second-order accuracy, the Monotonic Upstream-Centered Schemes for Conservation Laws~(MUSCL)~\cite {Tom2018} approach is employed, and Spatial gradients are calculated using the Green–Gauss method~\cite{BLAZEK-GreenGauss}. A time-marching scheme, incorporating Euler implicit time integration, is implemented to reach a steady-state solution. Finally, the resulting linearized system is solved via the flexible generalized minimum residual (FGMRES) method~\cite{Saad2003IterativeMethodsBook}, supported by an incomplete LU (ILU) precondition.

\subsection{Adjoint solver}
The adjoint solver provides sensitivity of the objective and the constraint values with respect to the design variables. To achieve this, the flow solver is differentiated using the discrete adjoint solver available within~\textit{SU2}. The implemented adjoint equations can then be derived by following the Lagrangian approach as elaborated in Ref.~\cite{SU2_INC}. Following this, the adjoint equation of the form
\begin{eqnarray}
    \label{eq:adjoint}
    \frac{d \mathcal{J}}{d\varepsilon}^\top = \frac{\partial \mathcal{J}}{\partial \varepsilon}^\top - \chi^\top \frac{\partial \mathcal{R}}{\partial \varepsilon},
\end{eqnarray}
 is obtained, where $\mathcal{J}$ is the objective function, $\mathcal{R}$ is the residual vector corresponding the primal solver and $\chi$ is the adjoint variable.

The source term corresponding to porosity, as shown in equation \eqref{eq:source_term}, is included in the residual term $\mathcal{R}$. {In addition, the porosity field ($\boldsymbol{\varepsilon}$) is registered as a dependent variable in the adjoint solver}. The sensitivity of the objective function with respect to the design variable is obtained using the algorithmic differentiation tool \textit{CoDiPack}~\cite{codipack}.
\section{Optimization Formulation}
\label{sec:topology_optim_probelm}
A generic topology optimization problem for fluid flow can be mathematically defined as
\begin{eqnarray}
\min_{\varepsilon} && \mathcal{J}(\boldsymbol{\varepsilon}), \\
\text { s.t. } && \boldsymbol{g}_i(\boldsymbol{\varepsilon}) \leq 0, \quad \forall i=1, \ldots, N \\
&& 0 \leq \boldsymbol{\varepsilon} \leq 1, 
\end{eqnarray}
where $\mathcal{J}$ is the objective function, while $\boldsymbol{\varepsilon}$ denotes the vector of design variables, representing the porosity value at each node in the flow domain. The vector $\boldsymbol{g}$ encapsulates a set of inequality constraints that the optimization problem must satisfy. During the design process, the optimization algorithm operates on $\boldsymbol{\varepsilon}$, which, as defined earlier, is a continuous variable ranging from $0$ for solid to $1$ for fluid, while the intermediate values represent porous materials. \review{Once the design converges, a threshold is applied to map these regions to a manufacturable solid/void geometry.}  The objective of the optimization process is to identify the optimal distribution of porosity that minimizes the objective function while satisfying the imposed constraints.

The choice of the objective function $\mathcal{J}$ and the constraints $\mathbf{g}$ is inherently linked to the specific problem considered. In the context of employing topology optimization for manifold design, the research reported in this manuscript focuses on three distinct optimization strategies. These approaches are, namely, the volume-only approach (VOA), the individual channel approach (ICA), and the flow uniformity-based approach (FUA). Among these, the first two formulations (VOA and ICA) are the most commonly discussed approaches in the literature, while the third formulation~(FUA) is a novel contribution of this study. The mathematical representation of the three formulations is presented in \autoref{tab:overview_formulations}, and a detailed description is reported in the following subsections. 

\begin{table}[t]
\centering
\caption{Overview of different manifold design optimization formulations.}
\label{tab:overview_formulations}
\resizebox{0.6\linewidth}{!}{%
\begin{tabular}{@{}llll@{}}
\toprule
 & \multicolumn{1}{c}{\begin{tabular}[c]{@{}c@{}} \Large \voa \end{tabular}} & \multicolumn{1}{c}{\begin{tabular}[c]{@{}c@{}} \Large \ica \end{tabular}} & \multicolumn{1}{c}{\begin{tabular}[c]{@{}c@{}} \Large \fua \end{tabular}} \\ \midrule
\begin{tabular}[c]{@{}l@{}}$\begin{array}{ll}\\ & \Large \text{min}_{\displaystyle \varepsilon}\\ \\ & \Large \text{s.t.}  \\ \\ \\ \\ & \Large \text{bounded by} \\ \end{array}$\end{tabular} & \begin{tabular}[c]{@{}l@{}}$\begin{array}{ll}\\ &\Large \mathcal{J} = \mathcal{P}(\boldsymbol{\varepsilon})\\ \\ & \Large g_{\text{vol}}(\boldsymbol{\varepsilon}) \leq 0 , \\ \\ \\ \\ & \Large 0 \leq \boldsymbol{\varepsilon} \leq 1\\ \end{array}$\end{tabular} & \begin{tabular}[c]{@{}l@{}}$\begin{array}{ll}\\ &\Large \mathcal{J} = \mathcal{P}(\boldsymbol{\varepsilon})\\ \\ &\Large g_{\text{vol}}(\boldsymbol{\varepsilon}) \leq 0 , \\\\ &\Large \boldsymbol{g}_{\dot{m}}(\boldsymbol{\varepsilon}) \leq 0 , \\\\ &\Large 0 \leq \boldsymbol{\varepsilon} \leq 1\\ \end{array}$\end{tabular} & \begin{tabular}[c]{@{}l@{}}$\begin{array}{ll}\\ &\Large \mathcal{J} = \mathcal{P}(\boldsymbol{\varepsilon})\\ \\ &\Large g_{\text{vol}}(\boldsymbol{\varepsilon}) \leq 0 , \\\\ &\Large g_{\phi}(\boldsymbol{\varepsilon}) \leq 0 , \\\\ &\Large 0 \leq \boldsymbol{\varepsilon} \leq 1\\ \end{array}$\end{tabular} \\ \bottomrule
\end{tabular}%
}
\end{table}

% \subsection{Traditional Optimization Strategy}
\subsection{Volume-Only Approach (VOA)}
The volume-only approach originates from the seminal work on density-based topology optimization for Stokes flow by Ref.~\cite{parallel_pipe}, which was subsequently extended to incompressible Navier-Stokes flow by Ref.~\cite{GersborgHansen2005TopologyOO}. In this approach, the objective function is defined based on the rate of energy dissipation and is defined as
\begin{equation}
\label{eq:powerdiss}
\mathcal{P} (\boldsymbol{\varepsilon}) = \int_{\Omega_\text{f}} ( 2 \mu \overline{\mathbf{S}}:\overline{\mathbf{S}} +  \rho \alpha (\varepsilon) \boldsymbol{v} \cdot \boldsymbol{v} ) \; d \Omega_\text{f},
\end{equation}
where, $\Omega_\mathrm{f}$ denotes the volumetric fluid domain and \review{$\overline{\boldsymbol{\mathrm{S}}}$ is the mean strain rate tensor defined as}
\begin{equation}
\overline{\mathbf{S}}=\frac{1}{2}\left(\nabla \boldsymbol{v} + \nabla \boldsymbol{v}^\top\right).
\end{equation}

This expression in equation~\eqref{eq:powerdiss} arises from the scalar multiplication of the momentum equations by the velocity vector, resulting in a volume-based power dissipation that also incorporates the effect of the Brinkman penalization term~\cite{Dilgen2018}. 

To ensure that the problem is well-posed, this formulation also includes a volume constraint, limiting the maximum allowable material within the design domain. This constraint can be written as
\begin{equation}
\label{eq:volcons}
g_{\text{vol}}(\boldsymbol{\boldsymbol{\varepsilon}})=\frac{\sum V_i \; \varepsilon_i}{f V} -1,
\end{equation}
where the notation $V_i$ stands for the volume of the $i$th cell and $\varepsilon_i$ represents its corresponding porosity value. Furthermore, $V$ represents the volume of the design domain, and $f$ denotes the target fluid volume fraction.

This \voa\ formulation is a widely adopted approach for fluid flow topology optimization, for example, in Ref.~\cite{MEN2024122008,  Dilgen2018, parallel_pipe, Feppon2020}.

% \subsection{Manifold-Specific Optimization Strategy}
\subsection{Individual Channel Approach (\ica)}
The \ica\ formulation builds upon the traditional \voa\ approach, by including adaptations specific to manifold design, as reported in Ref.~\cite{Kubo2016, Zeng2018}. In \ica, the objective function remains the same as that in \voa, that is, power dissipation ($\mathcal{P}$) as defined in equation~\eqref{eq:powerdiss}, while the constraints include additional mass flow rate restrictions for each outlet.
%
%additional constraints beyond the volume constraint ($g_{vol}$), particularly related to the mass flow rates at each of the outlets.
%Dont think we need this again
%To ensure a specified distribution of flow from the inlet to multiple outlets, the mass flow rate at each of the outlet channels from the manifold is constrained. 
These additional constraints are defined as
\begin{equation} 
\label{eq:g_MF}
\boldsymbol{g}_{\dot{m}}(\boldsymbol{V}(\boldsymbol{\varepsilon}))=\frac{\int_{A_i} \mathbf{n} \cdot \boldsymbol{v} \; d A}{\theta_{i} \;\dot{m}_\mathrm{in}} - 1 - \delta \;, \quad i=1, \dots, n \; 
\end{equation}
where $\dot{m}_\mathrm{in}$ is the total inlet mass flow rate, $\theta_{i}$ represents the desired fraction of inlet flow at the $i$-th channel, $n$ is the total number of channels over which the flow has to be distributed, and $\delta$ is the acceptable tolerance on the specified mass flow fractions.

As evident from equation~\eqref{eq:g_MF}, constraints $\mathbf{g}_{\text{MF}}$ depend on state variables, involving flux evaluations across cell faces. This dependency implies that, in an adjoint-based optimization framework, an additional adjoint equation must be solved for each outlet to compute the required sensitivities accurately. Hence, the number of adjoint evaluations required will scale with the number of outlets in this approach.

\subsection{Flow Uniformity based Approach (FUA)}
The \fua\ formulation, proposed in this study, builds upon the previously mentioned \voa\ formulations by proposing extensions specific to the manifold design problem. As with the earlier approaches, the objective function in \fua\ formulations remains as power dissipation ($\mathcal{P}$). However, an additional constraint, namely the flow uniformity coefficient, is introduced alongside the volume constraint in this formulation.

This flow uniformity coefficient ensures uniform flow distribution across the manifold's multiple outlets. This coefficient utilizes an ensemble quantity to represent flow uniformity, rather than restricting individual outlet flow rates (as used in \ica). This flow uniformity coefficient is defined as the standard deviation of mass flow rates across the outlets, as shown in equation~\eqref{eq:phi}. Thus, $\phi$ equals 0\% for a uniformly distributed flow and takes on a value greater than 0 in the presence of any flow maldistribution. In equation~\eqref{eq:phi}, $\dot{m}_i$ is the mass flow rate in the $i$-th channel, $\dot{m}_{\mathrm{avg}}$ is the average mass flow rate across all the channels and $n$ is the total number of channels.

\begin{equation} 
\label{eq:phi}
\phi = \sqrt{\frac{\sum_{i=1}^n\left(\frac{\dot{m}_i\; - \; \dot{m}_{\mathrm{avg}}}{\dot{m}_{\mathrm{avg}}}\right)^2}{n}} \times 100 \%
\end{equation}

In the \fua\ formulation, this coefficient $\phi$ is then used directly as a constraint in the design optimization process, expressed as
\begin{equation} \label{eq:g_phi}
g_{\phi}(\boldsymbol{V}(\boldsymbol{\varepsilon})) = \frac{\phi}{\phi_{\text{ref}}} - 1 \; \leq \; 0, \end{equation}
where $\phi_{\text{ref}}$ represents the desired value of the flow uniformity coefficient.

Unlike the \ica\ formulation, which constrains the mass flow rate at each outlet individually through $g_\text{MF}$, the \fua\ approach only requires a single constraint for flow uniformity, independent of the number of outlets. As a result, this formulation reduces the number of constraints in the optimization problem, aiming to achieve an even flow distribution with the optimized designs.

%Consequently, this formulation simplifies the optimization problem by limiting the number of constraints while still ensuring an even flow distribution
\subsection{Optimization Algorithm}
The entire topology optimization framework is implemented in an in-house Python-based optimization framework, called \textit{FlowForge}. This framework interfaces with the \textit{SU2} to obtain flow objectives and their constraints, and provides them to the employed optimization algorithm. To consistently communicate with different optimization algorithms, this framework integrates the open-source Python package \textit{pyOptSparse}~\cite{wu2020pyoptsparse}.
% % VERIFICATION
% \input{Tex/4a_Verification}
% % TEST CASE
\section{Case Study - Planar z-type Manifolds}
\label{sec:case_study}  
\begin{figure}[t]
\centering 
    \includegraphics[width=0.5\linewidth]{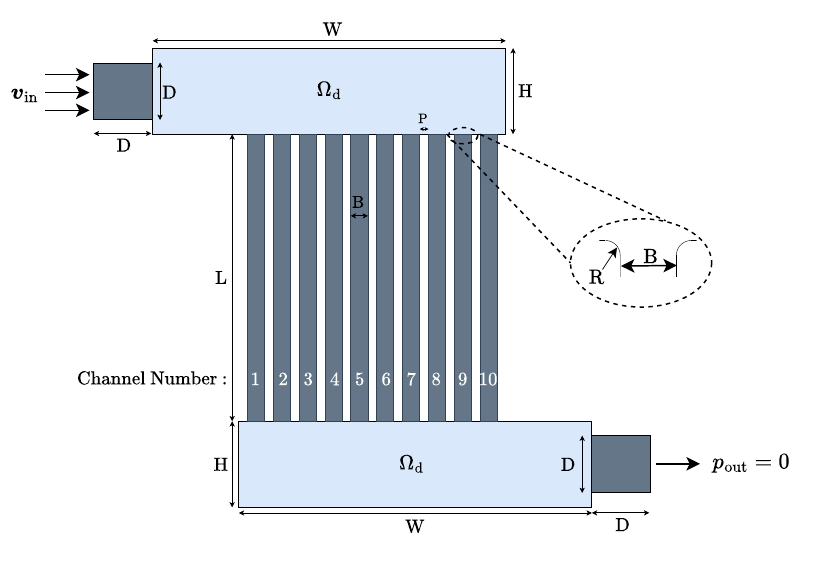}
    \caption{Geometric illustration of the planar z-type flow manifold case. The light blue regions indicate the design domains subject to optimization, while the dark gray regions represent fluid domains not participating in the optimization process.}
\label{fig:design_domain}
\end{figure}

To evaluate different optimization formulations, a planar z-type manifold with ten parallel channels was selected. Such channel configurations find application in micro-channel heat exchangers~\cite{app12136604}, reactors~\cite{Kubo2016}, and compact tube heat exchangers~\cite{KALLATH2021116645}. To achieve uniform flow distribution in such a parallel flow manifold is inherently challenging due to the orientation of the outlet channels relative to the inlet, as established in the literature by Ref.~\cite{ANBUMEENAKSHI2016156, XIA2015439}.

A schematic of the selected configuration is presented in~\autoref{fig:design_domain}. As illustrated, the inlet and the outlet manifolds form the design domain, denoted by $\Omega_{\text{d}}$ in~\autoref{fig:design_domain}, while the ten parallel channels do not participate in optimization. In the figure, the channel numbering is defined such that channel 1 is located closest to the inlet, while channel 10 is the farthest downstream. The geometrical values used for this case are detailed in~\autoref{tab:2D_case_geometry}.

\begin{table}[t]
\centering
\caption{Geometric parameters used in the planar z-type flow manifold case.}
\label{tab:2D_case_geometry}
\resizebox{0.55\linewidth}{!}{%
\begin{tabular}{@{}lccr@{}}
\toprule
\textbf{Parameter} & \textbf{Symbol} & \textbf{Unit} & \textbf{Value} \\ \midrule
manifold width & W & m & $61.5\times10^{-3}$ \\
manifold height & H & m & $15\times10^{-3}$ \\
inlet/outlet length & D & m & $10\times10^{-3}$ \\
channel width & B & m & $3\times10^{-3}$ \\
channel length & L & m & $50\times10^{-3}$ \\
channel pitch & P & m & $1.5\times10^{-3}$ \\
channel fillet radius & R & m & $0.25\times10^{-3}$ \\
number of channels & $n$ & - & $10$ \\ \bottomrule
\end{tabular}%
}
\end{table}

The flow in the selected case is simulated at an inlet velocity of 1.0~m/s, while the outlet is set to a gauge pressure of 0~Pa. Besides, no-slip boundary conditions are enforced along the manifold and channel walls. The thermo-physical properties of the fluid are set to constant, with a density of 995.7~$\text{kg}/\text{m}^{3}$ and dynamic viscosity of $9.975~\times10^{-3}$ Pa s. Thus, the flow in the manifold corresponds to a Reynolds number (Re) of 1000, given the inlet velocity and the inlet length as the characteristic length. 

The results reported in this manuscript utilize Sparse Nonlinear OPTimizer (SNOPT)~\cite{SNOPT}, a gradient-based optimization method, to solve the constrained optimization problem. The SNOPT algorithm updates the design variables based on the gradient information of both the objective function and constraints. \review{The optimality and feasibility tolerance value of $1 \times 10^{-6}$ were set as optimization termination criteria}. The optimization parameters considered for this case are summarized in \autoref{tab:optim_param}, and the number of design variables for this case is 30099.

\begin{table}[t]
\centering
\caption{Configuration parameters used for the topology optimization of flow manifolds.}
\label{tab:optim_param}
\resizebox{0.5\linewidth}{!}{%
\begin{tabular}{@{}lllll@{}}
\toprule
\multicolumn{1}{c}{$\alpha$} & \multicolumn{1}{c}{$q$} & \multicolumn{1}{c}{$f$} & \multicolumn{1}{c}{$\delta$} & \multicolumn{1}{c}{$\phi_{\text{ref}}$} \\ \midrule
$1 \times 10^{3}$ &  [$0.01$, $0.1$, $1$] & $0.6$ & $1 \times 10^{-3}$ & $1 \times 10^{-3}$ \\ \bottomrule
\end{tabular}%
}
\end{table}
% % Verification
% \input{Tex/4a_Verification}
% % RESULTS
\section{Results}
\subsection{Gradient Verification}
\begin{figure}[ht]
    \centering
    \includegraphics[width=0.7\linewidth]{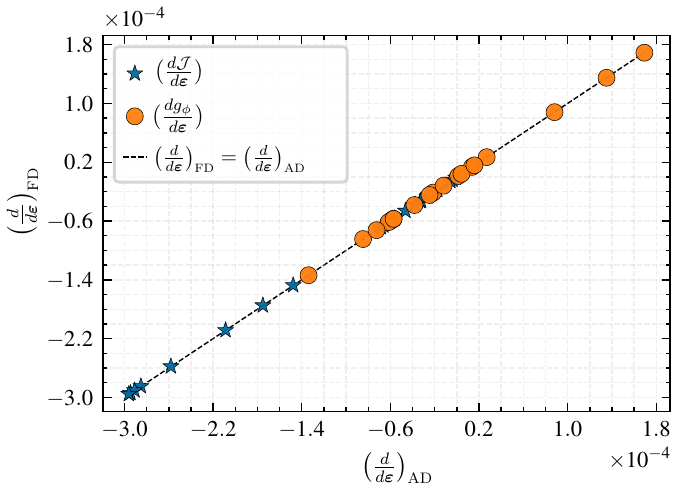}
    \caption{Comparison of adjoint gradients (AD) against finite-difference gradients (FD) obtained for the planar z-type flow manifold case.}
    \label{fig:gv_combi}
\end{figure}

To assess the accuracy of the gradient values obtained using the adjoint~(AD) method, the sensitivity values obtained are compared with those calculated using the finite difference (FD) method. \review{In the current study, a first-order accurate forward finite difference method with a step size of $1\times 10^{-5}$ was used. This step size was determined from a sweep of step sizes that minimized the finite-difference error, balancing truncation and round-off effects.} \autoref{fig:gv_combi} presents the comparison of the sensitivities obtained using the two methods, AD and FD, for the power dissipation~($\mathcal{J}$) and the flow uniformity~($g_{\phi}$). 

It can be observed that for 20 randomly selected design variables, the two sensitivities correlate well for the two cost functions, $\mathcal{J}$ and $g_{\phi}$. The average deviation between the sensitivities obtained from the two methods is less than 1\%. Hence, it can be concluded that the adjoint method was accurately implemented and the adjoint solver has sufficiently converged to provide accurate sensitivity values required for the optimization study.

\subsection{Comparison of Optimization Strategies}
The performance of the three optimization formulations described in \autoref{sec:topology_optim_probelm}, namely, \voa, \ica\ and \fua, are evaluated using the planar z-type manifold case, elaborated in \autoref{sec:case_study}. The optimizations were carried out on a high-performance computational infrastructure featuring 28 cores of an \textit{Intel Xeon Gold 5220R} (2.2 GHz) processor, complemented by 256 GB of memory.

\begin{figure}[t]
    \centering
    \includegraphics[width=0.7\linewidth]{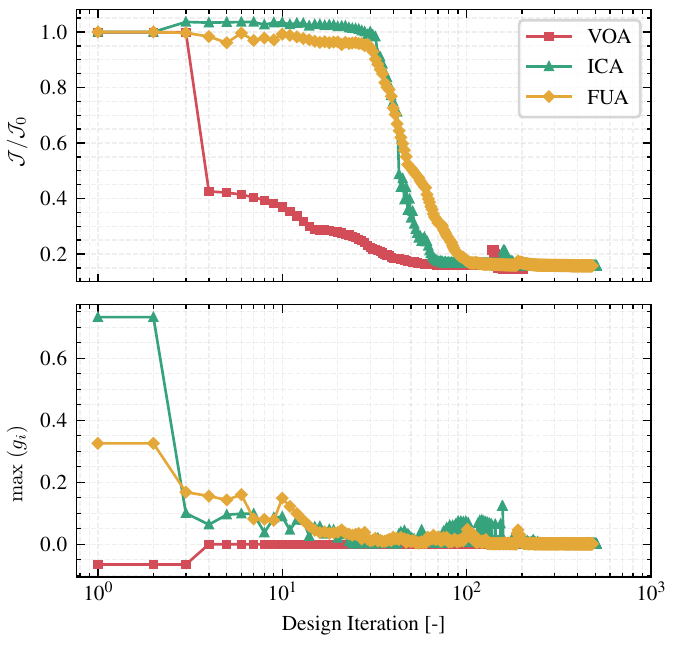}
    \caption{Optimization history objective function ${\mathcal{J}}/{\mathcal{J}_o}$ (top) and \review{maximum normalized constraint violation} $g_{i}$ (bottom).}
    \label{fig:opt_hist}
\end{figure}

To obtain the optimization results, the design domain was initialized with porosity values of 0.5. Optimal manifold designs were obtained in all three cases by substantially reducing the objective function while respecting the imposed constraints. \autoref{fig:opt_hist} presents the obtained optimization history for the three strategies. The top figure illustrates the evolution of the objective function, and \review{the bottom shows the variation of the maximum normalized constraint violation across all constraints during the optimization process (see~\autoref{fig:opt_hist})}. 

% Since the number of constraints differs between optimization formulations, the maximum value is shown to provide a consistent comparison.}
%

Based on the results obtained, \autoref{tab:optim_result} compares the key performance indicators of the three optimization strategies considered in this study. In this table, the performance of the baseline manifold configuration (initial geometry with all fluid nodes) is compared with the designs obtained using the three strategies. In terms of fluid dynamic performance, it can be observed that all three strategies have a higher value of the objective function when compared to that obtained for the baseline design. This is primarily due to the higher maximum velocity in the optimum manifolds. Although the optimum designs obtained using \ica\ and \fua\ exhibit comparable power dissipation values, they are higher than that obtained by the optimum design from \voa. In contrast, the flow maldistribution coefficient is significantly lower for the geometries obtained using the \ica\ and \fua\ strategies when compared with baseline and \voa. Besides, the $\phi$ values from \ica\ and \fua\ were also found to be comparable to each other. Thus, it can be concluded that the \voa\ approach cannot provide an efficient design of the manifold. Meanwhile, both \ica\ and \fua\ can achieve superior flow uniformity.

\begin{table}[t]
\centering
\caption{Comparison of optimization results obtained from different strategies. The baseline design represents a fluid-only solution of the initial geometry.}
\label{tab:optim_result}
\resizebox{0.75\linewidth}{!}{%
\begin{tabular}{llllll}
\hline
\multicolumn{1}{c}{} & \multicolumn{1}{c}{\textbf{Units}} & \multicolumn{1}{c}{\textbf{Baseline}} & \multicolumn{1}{c}{\textbf{VOA}} & \multicolumn{1}{c}{\textbf{ICA}} & \multicolumn{1}{c}{\textbf{FUA}} \\ \hline
$\mathcal{J}/\mathcal{J}_{0}$ & - & 0.131 & 0.145 & 0.157 & 0.158 \\
$\phi$ & \% & 35.4 & 38.8 & 0.089 & 0.099 \\
total iterations & - & - & 203 & 508 & 481 \\
time per iteration & min & - & 2 & 12 & 3 \\ \hline
\end{tabular}%
}
\end{table}

Besides fluid dynamic performance, \autoref{tab:optim_result} also compares the computational cost associated with the different strategies. It can be observed that \ica\ takes the most number of optimization steps~(503), arguably due to the complex formulation of the optimization problem with multiple constraints. Meanwhile, \voa\ takes the least number of optimization steps~(203); however, it fails to achieve an even flow distribution. Meanwhile, \fua\ takes 481 design steps to reach an optimum design. Further, it can be observed that the computational cost per design step is significantly higher for \ica~($\sim$12~{min}) than that for \fua~($\sim$3~{min}). This is because in the \ica\ strategy, each channel mass flow is individually constrained, thus requiring 1 primal evaluation and 11 adjoint evaluations (1 objective and 10 constraints) per design step. In contrast, the ensemble mass flow approach employed by \fua\ strategy dramatically reduces the cost per design step and consequently the overall temporal cost of optimization.

\begin{figure}[t]
    \centering
    \includegraphics[width=0.95\linewidth, trim=0 320 0 20]{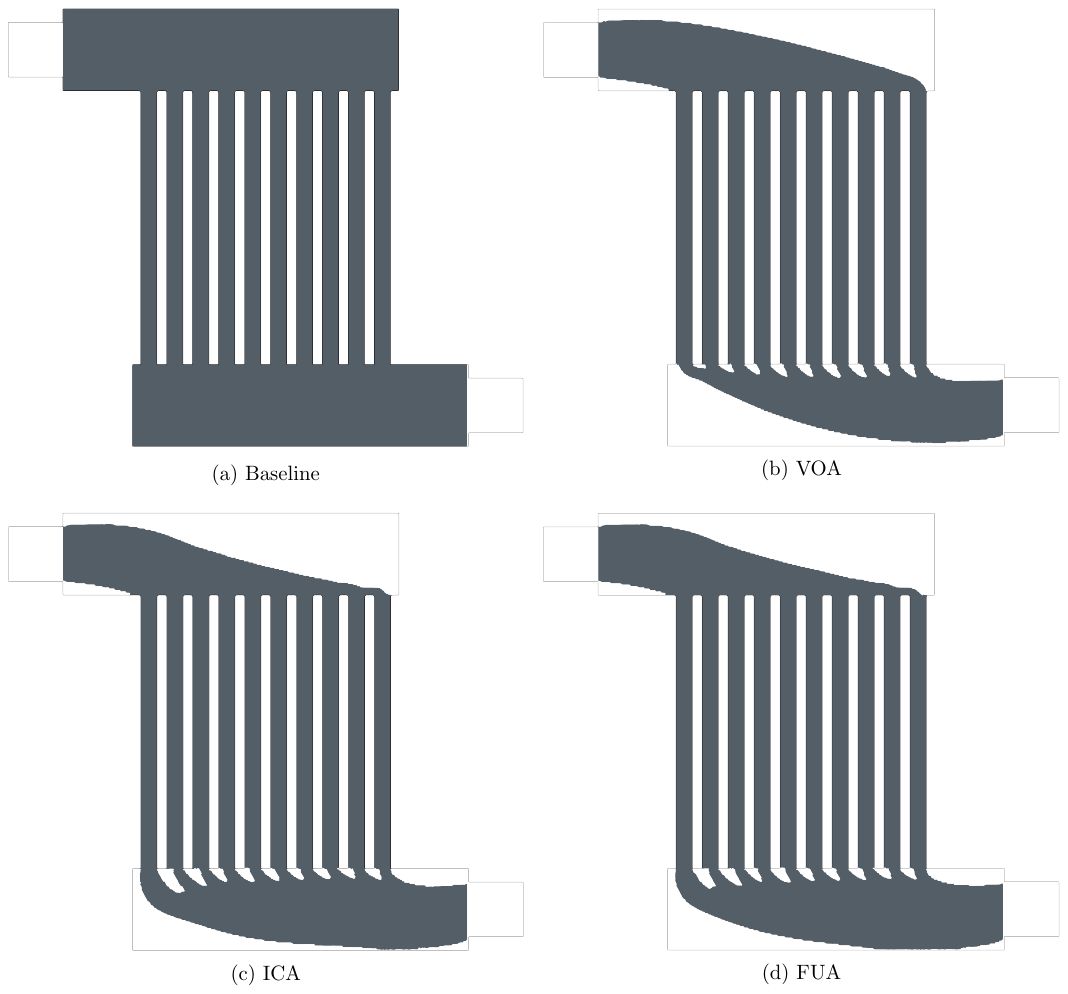}
    \caption{Comparison of the baseline design and optimal designs obtained through different optimization strategies for the planar z-type flow manifold.}
    \label{fig:design_vis_2}
\end{figure}

\autoref{fig:design_vis_2} presents the optimal manifold designs obtained for each optimization strategy. It can be observed that the geometries obtained using \ica\ and \fua\ strategies feature almost identical shapes. This is primarily due to the nearly perfect flow uniformity values obtained in the two optimal designs. In contrast, the shape obtained through \voa\ features noticeably different inlet and outlet manifold shapes. Specifically in the inlet manifold, \ica\ and \fua\ feature a sharp taper as compared to that obtained in \voa. In addition, it can be observed that the manifold connecting channels 7-10 features a narrow section as compared to that obtained through \voa, promoting smoother flow redirection. Meanwhile, in the outlet manifold, the transition from channels 1 to 5 is more gradual in \ica\ and \fua\, with rounded connections that help reduce sudden changes in flow direction.

\begin{figure}[t]
\centering
\includegraphics[width=0.6\linewidth]{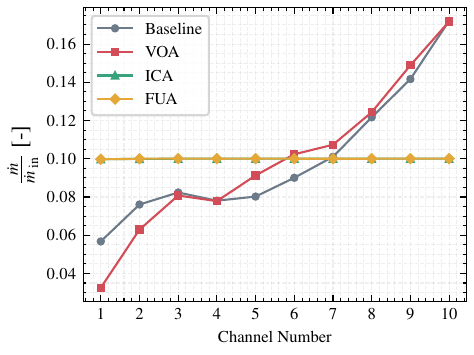}
\caption{Comparison of normalized mass flow rate variation across the channels for optimum designs obtained through different strategies, alongside the baseline design.}
\label{fig:mf_comparison}
\end{figure}

These geometric differences directly influence flow distribution in the channels, as shown in \autoref{fig:mf_comparison}. The baseline design exhibits significant flow non-uniformity, with channels farther from the inlet receiving disproportionately higher flow rates. This trend is consistent with observations reported in prior studies for a rectangular manifold design, for example, in Ref.~\cite{SIDDIQUI2017969}. Similarly, the optimum design obtained through \voa\ does not effectively balance the flow distribution, as channels near the inlet experience much lower flow rates than those farther downstream. In contrast, the optimum designs from \ica\ and \fua\ formulations achieve a near-uniform flow distribution. 

\begin{figure}[t]
    \centering
    \includegraphics[width=\linewidth]{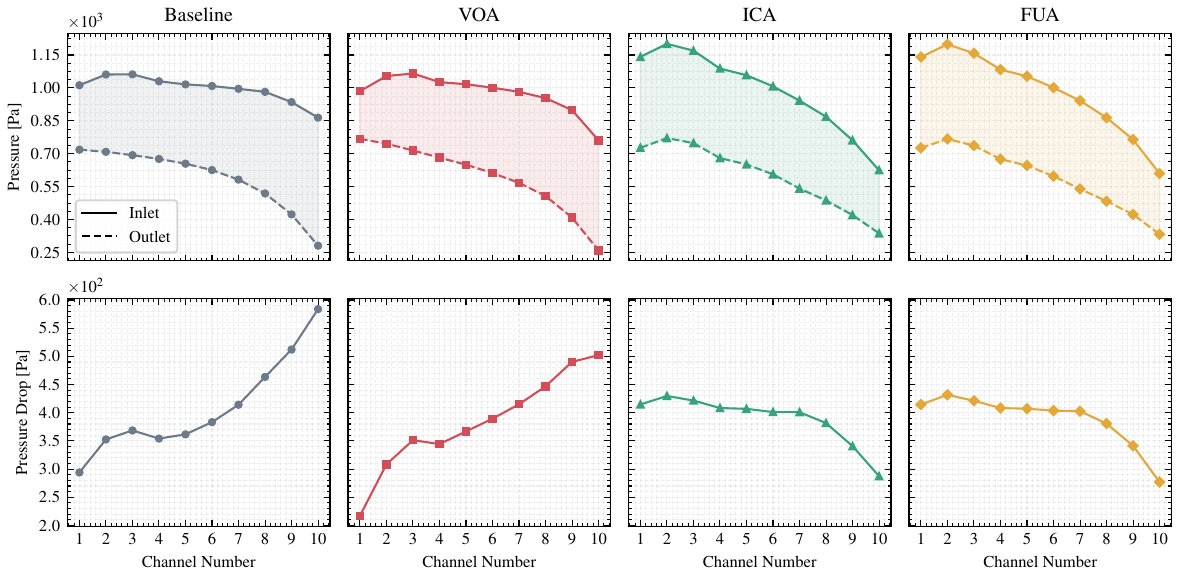}
    \caption{Comparison of channel-wise pressure variation (top) and pressure-drop (bottom) among different optimum designs obtained and the baseline design.}
    \label{fig:pressure_profiles}
\end{figure}

To further evaluate the flow distribution within the channels, \autoref{fig:pressure_profiles} presents the pressure at the inlet and outlet of each channel, along with their corresponding pressure drop values.  \autoref{fig:pressure_profiles} (top) illustrates the pressure profiles across the ten channels at both the inlet and outlet for each manifold design. It can be observed that, in the baseline and optimum design from \voa, the inlet pressure distribution is relatively uniform across all channels as compared to that obtained in optimum geometries from \ica\ and \fua. However, it can be seen that the outlet pressure features a steep gradient from channel 1 to channel 10, for baseline and the optimum geometry from \voa. This results in higher pressure drop values and, consequently, increased flow rates in channels 7 to 10, leading to a high flow maldistribution. In contrast, the designs obtained using \ica\ and \fua\ strategies display pressure variations from channel 1 to 10, at both the inlet and outlet. Although the variations seem comparable to those in \voa\ and the baseline case, the slope of the two pressure profiles (inlet and outlet) is similar. Hence, these pressure variations effectively compensate for each other, resulting in a more uniform pressure drop across the channels, as shown in \autoref{fig:pressure_profiles} (bottom). More specifically, the pressure drop remains nearly constant from channels 1 to 7, with a modest decline observed in the final channels. This leads to a more uniform flow distribution and thus a reduced maldistribution.

\begin{figure}[ht!]
\centering 
    \includegraphics[width=\linewidth, trim= 40 320 0 100]{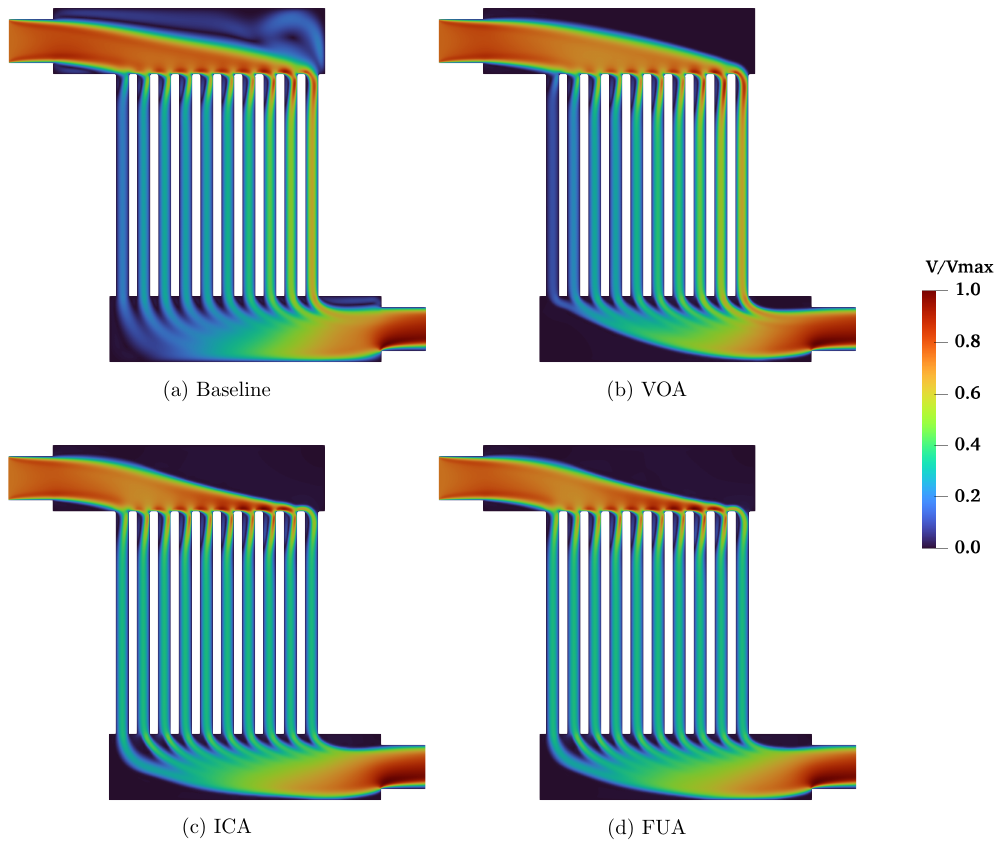}
    \caption{Mid-plane normalized velocity fields of optimum designs, alongside the baseline configuration.}
\label{fig:velocity_field_vis}
\end{figure}

\autoref{fig:velocity_field_vis} illustrates the velocity flow field obtained from CFD simulations for the four manifold configurations. It can be observed that the flow in the baseline design suffers from a large recirculation zone in the inlet manifold, particularly near channels 8-10. Additionally, flow separation can be observed in the baseline configuration at the outlet manifold, particularly near channels 1 through 4. In contrast, the optimized geometries feature reduced recirculation and flow separation zones by incorporating tapered inlet and outlet manifolds. Although beneficial, this tapered shape leads to sharp curvature, particularly near the channel junctions, which leads to steep velocity gradients. Consequently, leading to elevated velocities which increase the kinetic energy of the flow and, thereby, raise the overall power dissipation compared to the baseline geometry, as reported in \autoref{tab:optim_result}. This effect is especially pronounced in the \ica\ and \fua\ configurations, where the sharper transitions near the downstream channels (particularly channels 7 to 10) cause substantial local velocity amplification, directly contributing to higher energy losses in those regions.

The similarity in geometry and flow behavior between the \ica\ and \fua\ designs indicates that a global flow uniformity constraint alone is sufficient to achieve a balanced flow distribution. Unlike methods that impose individual mass flow rate constraints on each channel, the single-constraint formulation used in \fua\ simplifies the optimization process and significantly reduces computational cost. Although the \fua\ strategy is attractive as an alternative to the \ica\ strategy to obtain uniform flow manifolds, it must be noted that it cannot accommodate control over individual channel flow rates, as it directly uses an ensemble quantity.
\section{Application to three-dimensional cases}
To evaluate the scalability of the proposed method, \fua\ strategy is applied to three-dimensional cases. More specifically, it is applied to two commonly used manifold configurations: the cylindrical z-type flow manifold and the radial flow manifold. For both cases, the optimization is formulated as summarized in \autoref{tab:overview_formulations}, with simultaneous optimization of both the inlet and outlet manifolds.

\subsection{Cylindrical z-type flow manifolds}
\begin{figure}[t]
\centering
\includegraphics[width=0.5\linewidth]{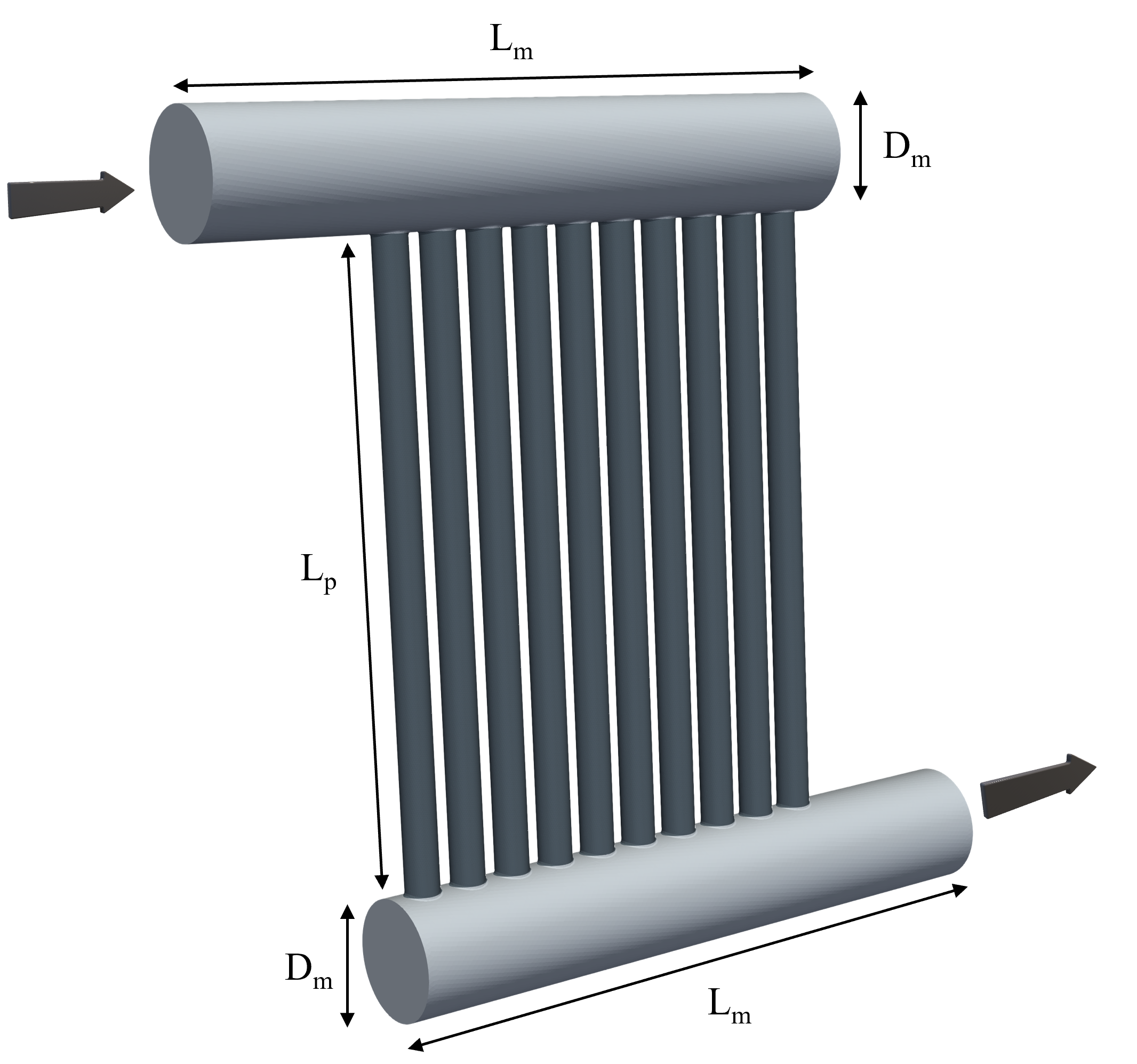}
\caption{Illustration of the geometry used for the cylindrical z-type flow manifold design case. The light blue volumes (manifolds) correspond to the design domains subject to optimization, whereas the dark gray volumes (channels) represent fixed fluid regions excluded from the optimization process.}
\label{fig:3D_design_domain}
\end{figure}

This test case directly extends the planar z-type flow manifold case discussed in \autoref{sec:case_study}. The baseline design in this case consists of cylindrical manifold geometries, as illustrated in \autoref{fig:3D_design_domain}. The channel numbering is the same as that used for the planar case, as shown in \autoref{fig:design_domain} and the geometric parameters used are listed in \autoref{tab:3D_parallel_manif_case_geometry}. The geometry features a fillet of radius 0.25 mm to prevent flow from turning over sharp corners. 

\begin{table}[t!]
\centering
\caption{Geometric parameters used in the cylindrical z-type flow manifold design case.}
\label{tab:3D_parallel_manif_case_geometry}
\resizebox{0.55\linewidth}{!}{%
\begin{tabular}{@{}llcr@{}}
\toprule
\textbf{Parameter} & \multicolumn{1}{c}{\textbf{Symbol}} & \textbf{Unit} & \textbf{Value} \\ \midrule
manifold diameter & $\text{D}_{\text{m}}$ & m & $15\times10^{-3}$ \\
manifold length & $\text{L}_{\text{m}}$ & m & $61.5\times10^{-3}$ \\
channel diameter & $\text{D}_{\text{p}}$ & m & $3\times10^{-3}$ \\
channel length & $\text{L}_{\text{p}}$ & m & $50\times10^{-3}$ \\
channel pitch & $s$ & m & $1.5\times10^{-3}$ \\
number of channels & $n$ & - & $10$ \\ \bottomrule
\end{tabular}%
}
\end{table}

 The flow in the cylindrical manifold case is simulated by imposing an inlet velocity of 1.0~m/s and a gauge pressure of 0~Pa at the outlet. Besides, no-slip boundary conditions are applied along the walls of both the manifolds and channels. The fluid is assumed to have constant thermo-physical properties, with a density of 995.7 $\text{kg}/\text{m}^{3}$ and dynamic viscosity of $9.975 \times10^{-3}$~Pa s. Under these conditions, the Reynolds number (Re) is approximately 1000, considering the inlet diameter as the characteristic length and the inlet velocity as the reference velocity. 
 
 The optimization problem was configured with parameters summarized in \autoref{tab:optim_param}, and the total number of design variables used was $\sim 250\times10^{3}$. As for the \twodim\ case, the inlet and outlet manifolds constitute the design domain, while the parallel channels are kept fixed during optimization. The optimization was initialized with a uniform porosity field of 0.6 and converged to a feasible design after 523 steps. The total computational cost was approximately 96 hours using 140 cores of \textit{Intel Xeon E5-2680v4} processors.

\begin{figure}[t!]
  \centering
    \centering
    \includegraphics[height=0.5\linewidth]{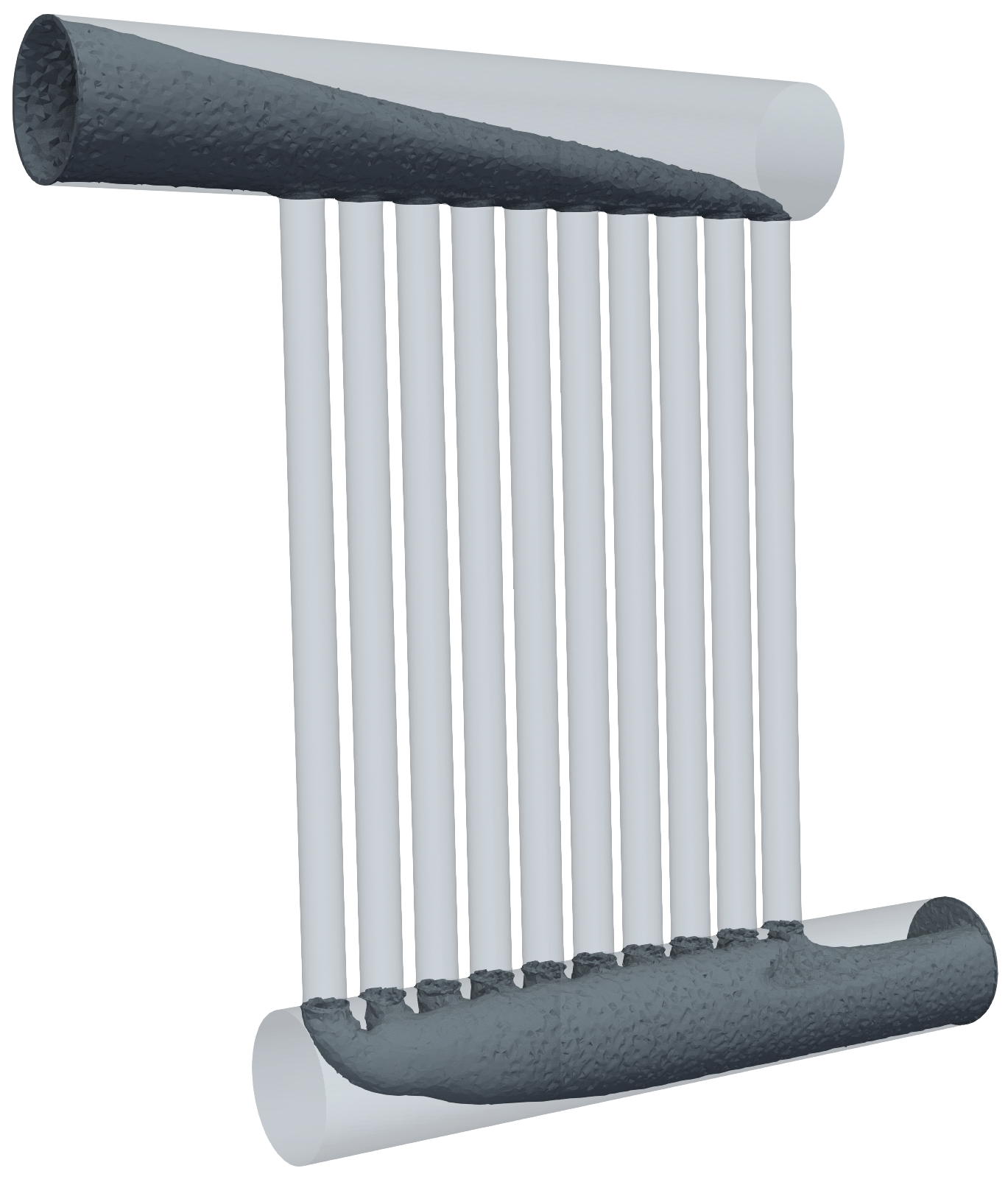}
    \caption{Optimum geometry obtained for the cylindrical z-type manifold case using the FUA strategy, highlighting the inlet and outlet manifolds.}
    \label{fig:3D_parallel_design}
\end{figure}

\autoref{fig:3D_parallel_design} depicts the optimum geometry obtained using the \fua\ optimization strategy for the cylindrical z-type manifold case. Consistent with the observation from the planar case, the inlet manifold exhibits a tapered configuration, whereas the outlet manifold is characterized by branched connecting structures.

\begin{figure}[ht!]
    \centering
    \includegraphics[width=0.8\linewidth]{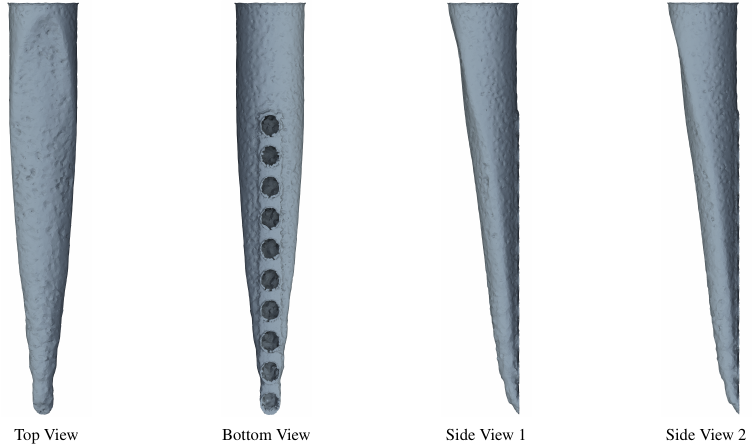}
    \caption{Orthographic views of the optimized inlet manifold for the cylindrical z-type flow manifold case, shown from the top, bottom, and two lateral perspectives.}
    \label{fig:3D_parallel_design_inlet_closeup}
\end{figure}

To better understand the geometrical features, an orthographic view of the inlet manifold is illustrated in \autoref{fig:3D_parallel_design_inlet_closeup}, which includes top, bottom, and side views. It can be observed that the manifold features a pronounced tapering along the primary flow direction, progressively narrowing to facilitate uniform flow distribution across the channels. Additionally, it can be observed that this tapering is symmetric about the midplane of the channel assembly.

\begin{figure}[ht!]
    \centering
    \includegraphics[width=0.8\linewidth]{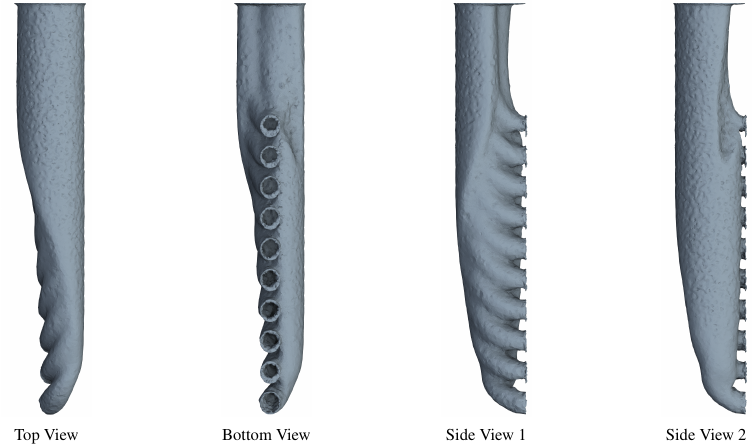}
\caption{Orthographic views of the optimized outlet manifold for the cylindrical z-type flow manifold case, shown from the top, bottom, and two lateral perspectives.}
\label{fig:3D_parallel_design_outlet_closeup}
\end{figure}

In contrast, the outlet manifold geometry, shown in \autoref{fig:3D_parallel_design_outlet_closeup}, is notably more complex and asymmetric. A close inspection reveals that each channel exit is connected to the outlet via distinct, curved flow paths that remain spatially distinct before converging into a single channel near the outlet. The manifold exhibits a swept-back shape, indicating a controlled and gradual redirection of flow from the channels into the outlet plenum.

\begin{figure}[t!]
    \centering
    \includegraphics[width=\linewidth]{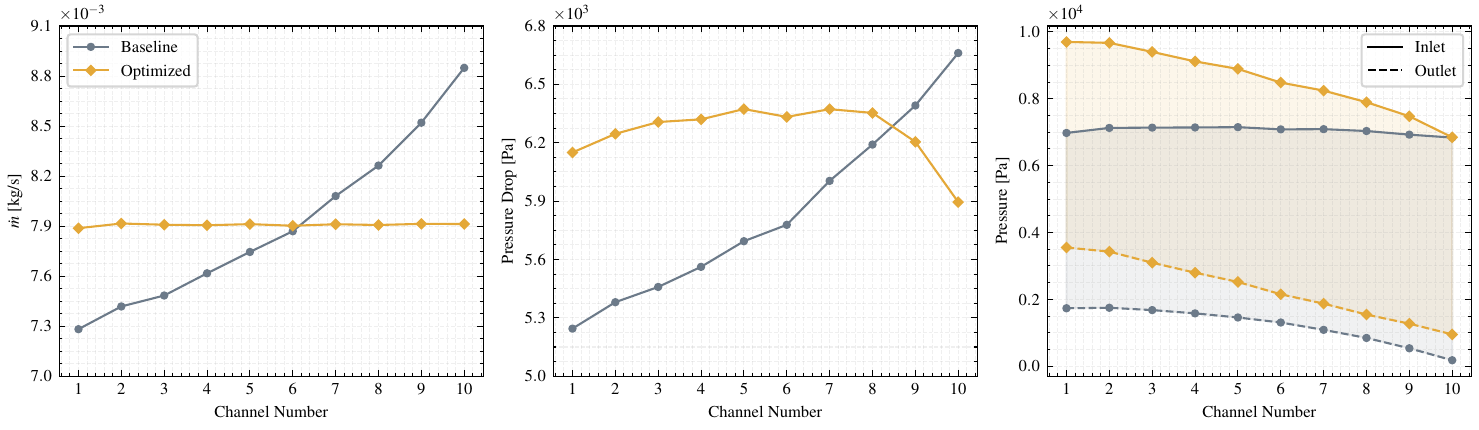}
\caption{Channel-wise comparison of flow properties, for the cylindrical z-type flow manifold case, between the baseline and the optimum configurations; mass flow rate (left), pressure drop (center), and pressure variation at the inlet and outlet (right).}
\label{fig:combi_comparison_3D_parallel}
\end{figure}

\autoref{fig:combi_comparison_3D_parallel} (left) compares the mass flow rate (\massflow) distribution obtained from the optimized and baseline designs across the 10 channels. The baseline design exhibits a notable imbalance in flow distribution; more specifically, the channels away from the inlet receive disproportionately high flow and vice versa. In contrast, the optimized geometry achieves highly uniform \massflow, with the flow uniformity coefficient ($\phi$) value of 0.1\%. This effect is directly attributable to the tapered inlet shape, which moderates flow acceleration and distributes it evenly across the channels. Besides, the segregated outlet geometry preserves individual channel flow paths and prevents cross-channel interference.

The impact of the optimized geometry on flow behavior is also illustrated in \autoref{fig:combi_comparison_3D_parallel}, which compares the inlet and outlet pressure distributions (right) and pressure drops (center) across the 10 parallel channels for both the baseline and optimized designs. The baseline design exhibits a relatively uniform inlet pressure, but a sharp decline in outlet pressure from channel 1 to channel 10. This consequently results in a steep gradient in pressure drop values and poor flow balance. This indicates limited outlet connectivity and unstructured flow coalescence. The optimized manifold, however, maintains a consistent pressure drop across all channels. The pressure drop curve is flatter, indicating a more balanced flow distribution and reduced variation in channel-level resistance. The tapering inlet and individually guided outlet corridors together contribute to balanced hydraulic resistance and reduced pressure variability between channels.

% \begin{figure}[h!]
% \centering 
%     \includegraphics[width=0.9\linewidth]{Images/3D_Case/Pressure_profile_comparison_v2.pdf}
%     \caption{Channel-wise inlet and outlet pressure distribution (top) and corresponding pressure drop (bottom) for the 3D cylindrical z-type flow manifold case.}
% \label{fig:pressure_profile_3D_parallel}
% \end{figure}

\begin{figure}[ht!]
\centering 
    \includegraphics[width=\linewidth]{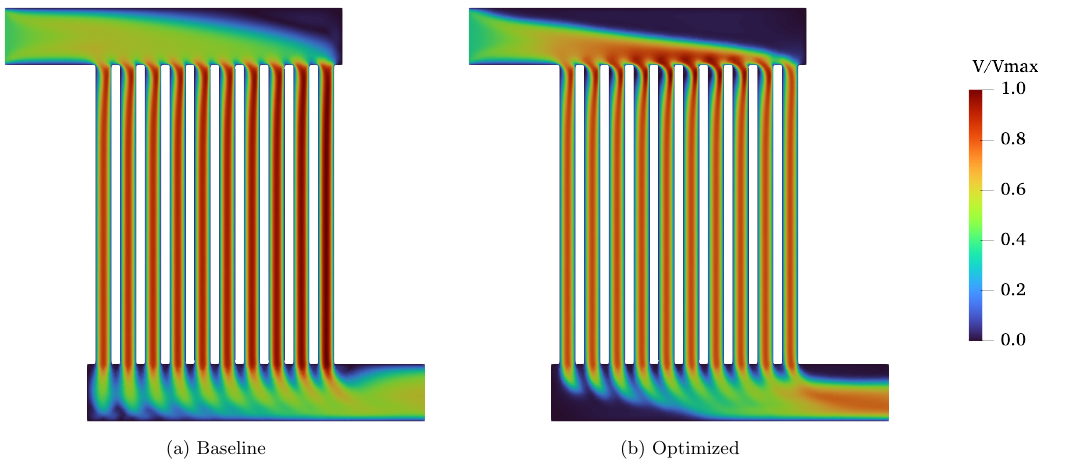}
    \caption{Mid-plane, normalized velocity fields of the cylindrical z-type flow manifold case; (a) baseline configuration, (b) optimized design.}
\label{fig:vel_mid_plane_3D_parallel_Case}
\end{figure}

\autoref{fig:vel_mid_plane_3D_parallel_Case} reports the mid-plane velocity magnitude contours normalized by the maximum velocity for baseline and the optimum designs. The baseline design (left) shows pronounced channel-to-channel velocity variation, with channels 6-10 having relatively high velocities compared to channels 1-5. In contrast, the optimized geometry (right) exhibits a much more uniform velocity field, consistent with the observed improvements in both mass flow rate distribution and pressure drop reported earlier. Nevertheless, similar to the planar case, the region of high velocity is increased in the optimized design, indicating a higher power dissipation compared to the baseline design.

\subsection{Radial manifold}
The next three-dimensional case focuses on manifolds commonly employed in chemical reactors and vertical heat exchangers~\cite{maldistriub_chemical, radial_case_Ref_1, radial_case_Ref_2}. In such manifolds, the fluid enters from the top and is radially distributed into multiple parallel channels through a cylindrical inlet manifold. Meanwhile, at the downstream end, a similar cylindrical manifold collects the flow from these channels and directs it toward the outlet pipe.

\begin{figure}[ht!]
\begin{subfigure}[t]{0.48\textwidth}
\centering
\includegraphics[height=0.8\linewidth]{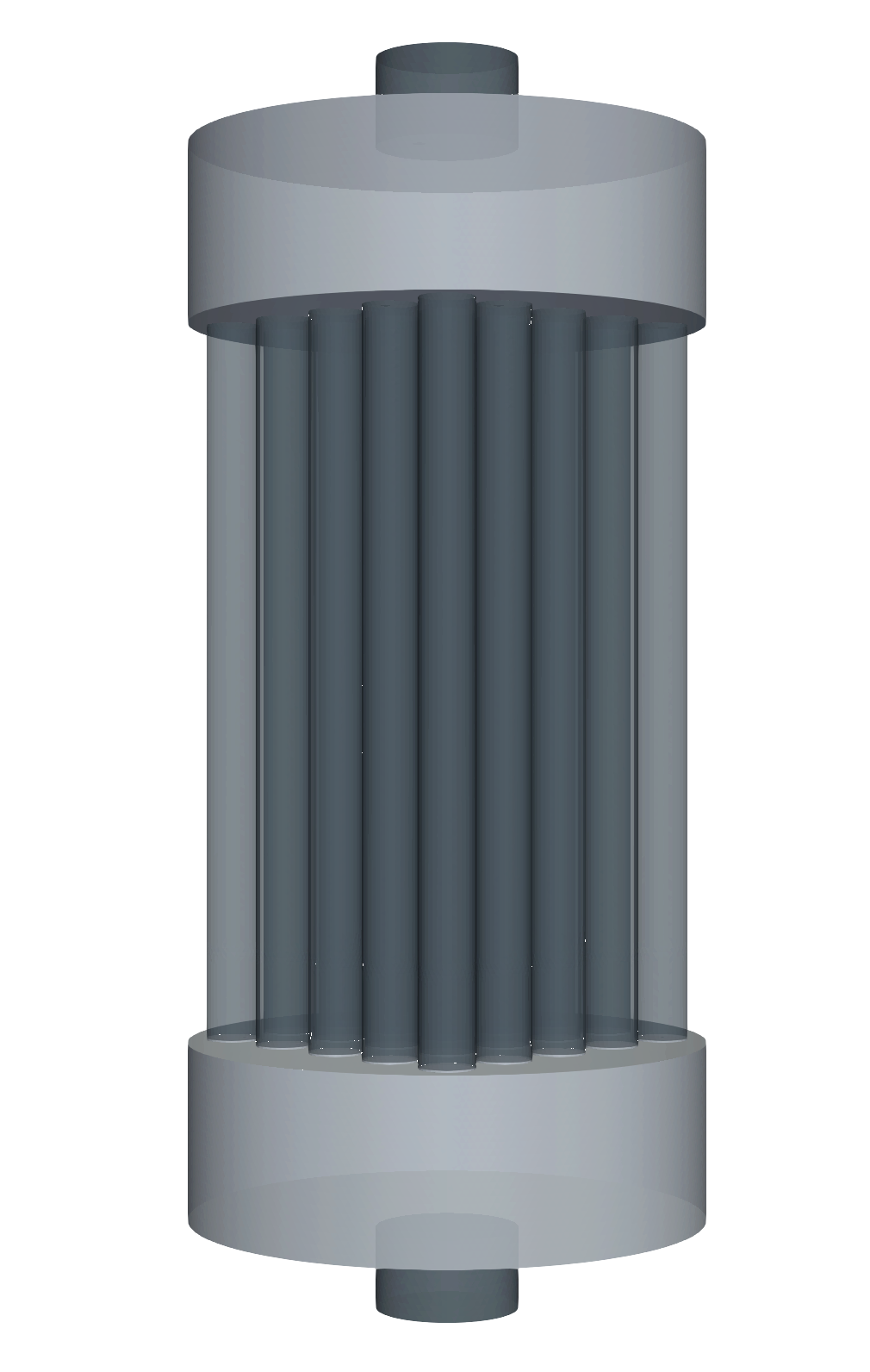}
\caption{Isometric view of the geometry used, where the light blue regions (manifolds) represent the design domains, and the dark gray regions (channels) represent fluid regions not participating in the optimization process.}
\label{fig:3D_circ_mainf_domain}
\end{subfigure}
\hspace{1em}
\begin{subfigure}[t]{0.48\textwidth}
\centering
\includegraphics[width=\linewidth]{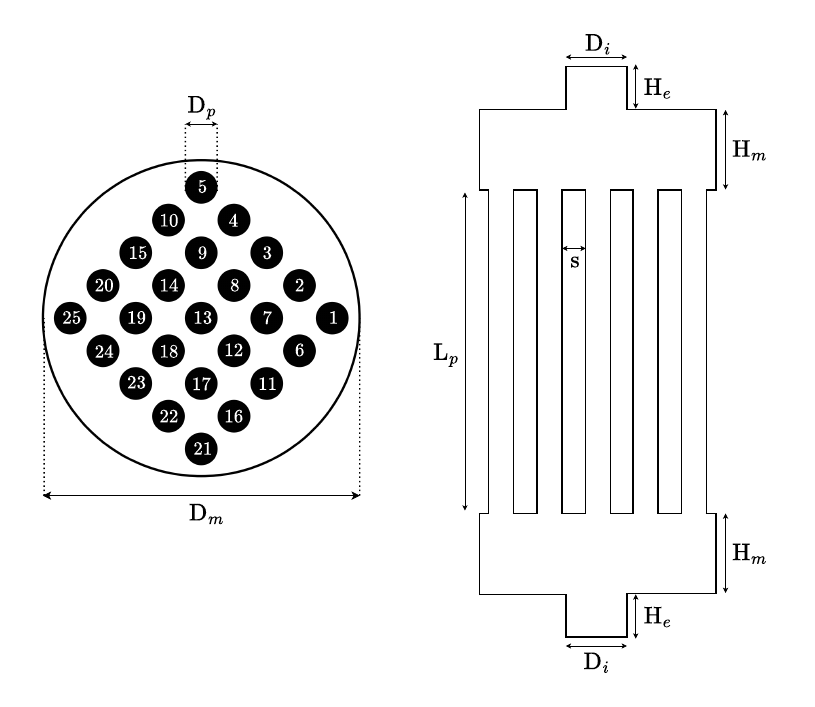}
\caption{Cross-sectional view of the radial manifold case, highlighting geometrical parameters. All the channels are numbered to enable discussion.}
\label{fig:3D_circ_mainf_domain_2}
% \vspace{1em}
\end{subfigure}
\caption{Illustration of the geometry used for the three-dimensional radial manifold case.}
\label{fig:radial_flow_3d_view}
\end{figure}

To optimize the design of such manifolds, an exemplary manifold geometry and topological configuration were selected and are illustrated in~\autoref{fig:radial_flow_3d_view}(a). The chosen geometry consists of two radial manifolds (inlet and outlet) encompassing a 5$\times$5 matrix of cylindrical channels. Cross-sectional views of the system, along with the geometrical specifications, are shown in \autoref{fig:radial_flow_3d_view}(b). The values of the geometrical specifications are tabulated in \autoref{tab:3D_circi_manif_case_geometry}. Each channel is numbered in the \autoref{fig:radial_flow_3d_view}(b) (circles) for reference in the analysis.

The flow in the selected case is simulated by imposing an inlet velocity of 1.25~m/s and a gauge pressure of 0~Pa at the outlet. Besides, no-slip boundary conditions are applied along the walls of both the manifolds and channels. The fluid is assumed to have constant thermo-physical properties, with a density of 995.7~$\text{kg}/\text{m}^{3}$ and dynamic viscosity of $9.975\times10^{-3}$ Pa s. Under these conditions, the Reynolds number (Re) is approximately 1000, given the inlet diameter as the characteristic length and the inlet velocity as the reference velocity. 

The optimization parameters used in this case are identical to those employed in previous cases, as summarized in \autoref{tab:optim_param}. The total number of design variables in this case is $\sim460\times10^{3}$. Starting from a uniform porosity field of 0.6, the optimization converged to a feasible solution in 732 iterations. The total computational time was approximately 142 hours on 168 \textit{Intel Xeon E5-2680v4} CPU cores. 

\begin{table}[ht]
\centering
\caption{Geometric parameters used in the radial manifold design case.}
\label{tab:3D_circi_manif_case_geometry}
\resizebox{0.55\linewidth}{!}{%
\begin{tabular}{@{}llcr@{}}
\toprule
\textbf{Parameter} & \multicolumn{1}{c}{\textbf{Symbol}} & \textbf{Unit} & \textbf{Value} \\ \midrule
manifold diameter & $\text{D}_{\text{m}}$ & m & $29\times10^{-3}$ \\
manifold height & $\text{H}_{\text{m}}$ & m & $10\times10^{-3}$ \\
inlet/outlet length & $\text{H}_{\text{e}}$ & m & $5\times10^{-3}$ \\
inlet/outlet diameter & $\text{D}_{\text{i}}$ & m & $8\times10^{-3}$ \\
channel diameter & $\text{D}_{\text{p}}$ & m & $3\times10^{-3}$ \\
channel length & $\text{L}_{\text{p}}$ & m & $40\times10^{-3}$ \\
channel pitch & $s$ & m & $3\times10^{-3}$ \\
number of channels & $n$ & - & $25$ \\ \bottomrule
\end{tabular}%
}
\end{table}

\begin{figure}[ht!]
  \centering
    \includegraphics[height=0.5\linewidth]{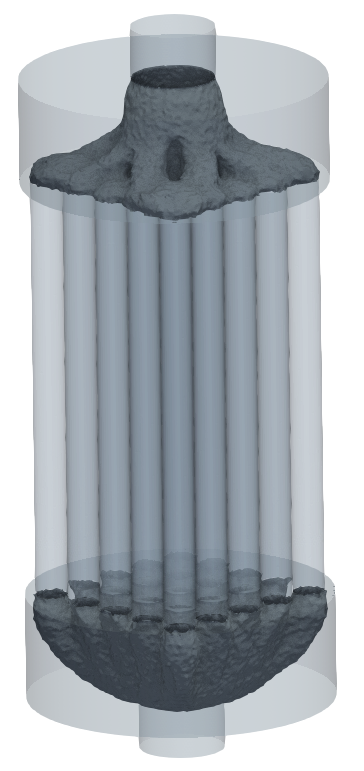}
    \caption{Optimal geometry obtained for the radial manifold case using the FUA strategy, highlighting the inlet and outlet manifolds.}
    \label{fig:3D_circ_design}
\end{figure}

\begin{figure}[ht!]
  \begin{subfigure}[t]{0.52\textwidth}
    \centering
    \includegraphics[width=\linewidth]{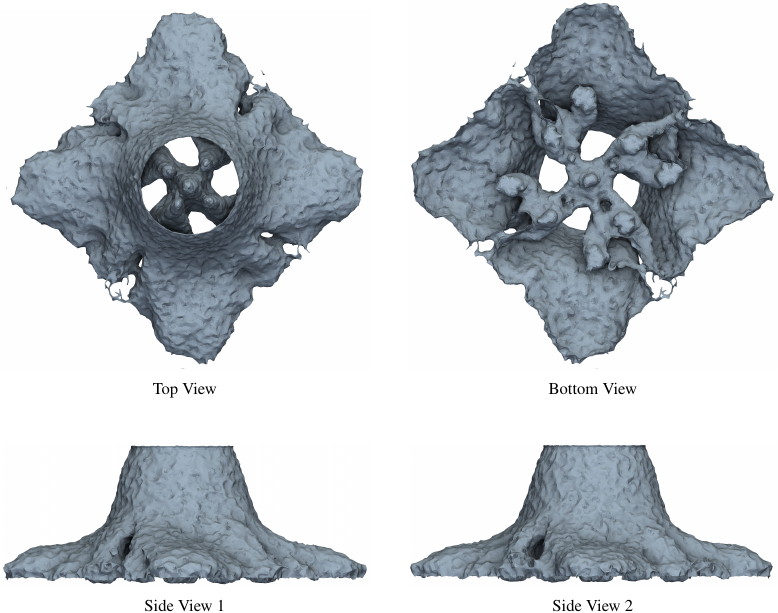}
    \caption{Orthographic view of the complete inlet manifold geometry, showing the top, bottom, and two side perspectives.}
    \label{fig:3D_circ_design_inlet_closeup}
  \end{subfigure}
  \hfill
  \begin{subfigure}[t]{0.44\textwidth}
    \centering
    \includegraphics[width=0.8\linewidth]{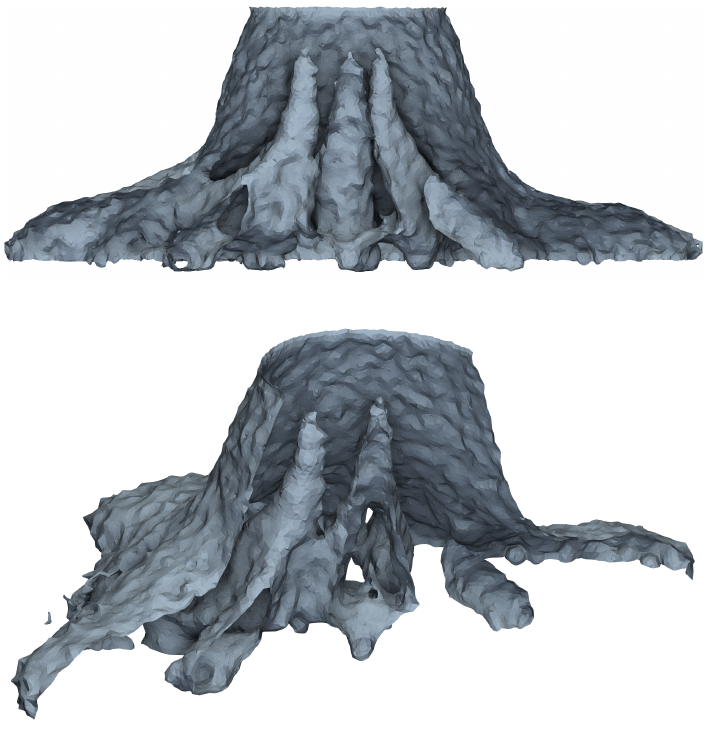}
    \caption{Sectional visualization of the inlet manifold, presented from a front view and an angled orientation.}
    \label{fig:3D_circ_design_inlet_cutaway}
  \end{subfigure}
  \caption{Visualization of the optimized inlet manifold obtained for the radial manifold case. (a) External views from multiple orientations. (b) Sectional views of the manifold.}
  \label{fig:combi_view_close_up_inlet_figures}
\end{figure}

\autoref{fig:3D_circ_design} shows the optimum geometry obtained using the \fua\ optimization strategy. A detailed view of the optimized inlet region is provided in \autoref{fig:combi_view_close_up_inlet_figures}(a), illustrating the \review{smoothly tapered transition} into the channel array. Internally, the optimized geometry features complex flow-dividing structures, as shown in \autoref{fig:combi_view_close_up_inlet_figures}(b). These features subdivide the inlet volume into multiple segments, introducing directional control within the plenum and shaping the internal flow path without requiring manual prescription. In addition, they also contribute to imparting rotational momentum to the flow, which promotes effective distribution toward the peripheral channels (e.g., 1, 5, 21, 25). 

\begin{figure}[htb!]
    \centering
    \includegraphics[width=0.55\linewidth]{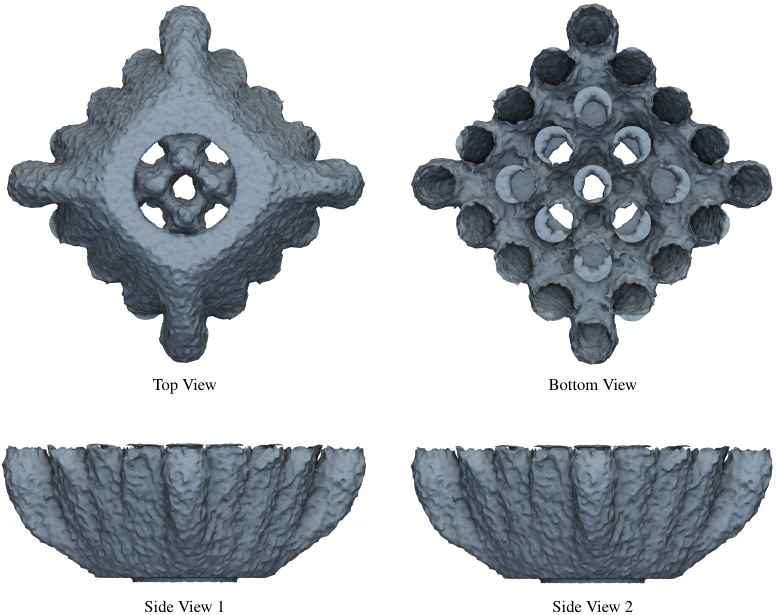}
    \caption{Orthographic views of the optimized outlet manifold geometry obtained for the radial manifold case, shown from the top, bottom, and two lateral perspectives.}
    \label{fig:3D_circ_design_outlet_closeup}
\end{figure}

\autoref{fig:3D_circ_design_outlet_closeup} shows the outlet manifold, which contrasts with the inlet manifold in structure. Instead of a symmetric taper, the outlet is characterized by a branching network that merges individual channel flows into a single exit port. The branches are smoothly contoured and spatially distributed to match the positioning of the channels while guiding the flow toward the outlet with minimal abrupt transitions.

\begin{figure}[ht!]
\centering
\includegraphics[width=\linewidth]{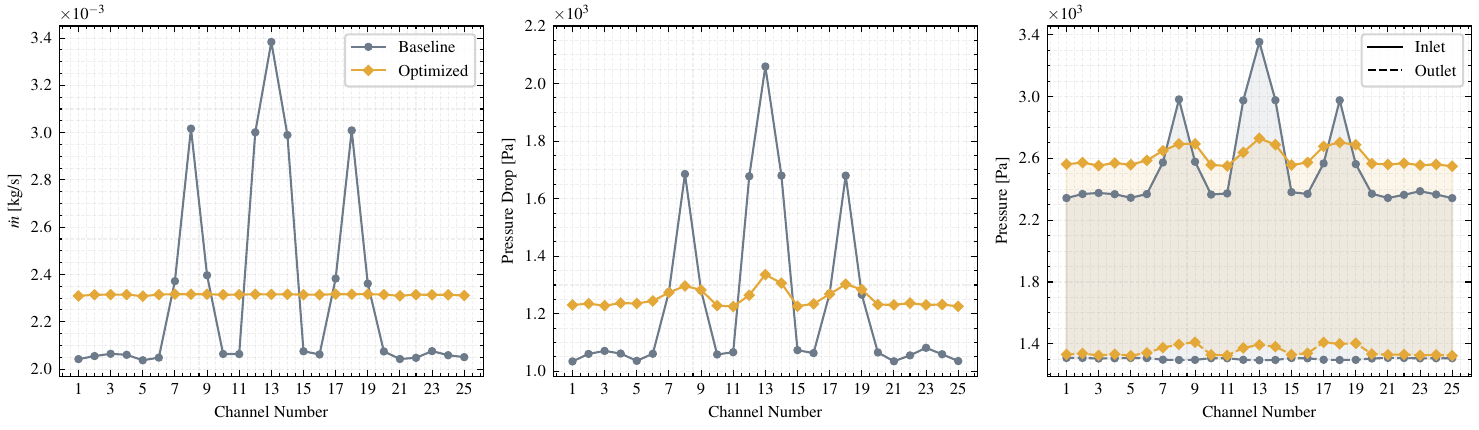}
\caption{Channel-wise comparison of flow properties between the FUA-optimized optimum geometry and baseline designs for the radial manifold case; mass flow rate (left), pressure drop (center), and pressure variation at the inlet and outlet (right).}
\label{fig:combi_comparison_3D_circ}
\end{figure}

\autoref{fig:combi_comparison_3D_circ} (left) shows the mass flow rate distribution across the 25 channels for both the baseline and optimized designs. In the baseline case with simple cylindrical manifolds, the flow is highly uneven, with much greater flow through the central channels due to a minimal resistance path. In contrast, the optimized design achieves near-uniform flow distribution. This improvement directly reflects the effects of the optimized inlet and outlet geometries: the radially tapering inlet, combined with the internal divider structures, ensures balanced flow distribution to all channels.

These trends are further explained by the pressure profiles shown in \autoref{fig:combi_comparison_3D_circ} (center, right). The baseline design exhibits significant variation in both inlet and outlet pressures, resulting in inconsistent pressure drops across the array. These imbalances drive the flow maldistribution observed earlier. The optimized design, on the other hand, shows flattened pressure profiles at both the inlet and outlet planes. The consistent pressure drops across channels confirm that the internal dividing features and branching outlet paths effectively homogenize resistance and eliminate preferential flow paths, resulting in a well-balanced flow distribution across the parallel channel array.

% \begin{figure}[h!]
% \centering 
%     \includegraphics[width=0.9\linewidth]{Images/3D_circ_case/Pressure_profile_comparison.pdf}
%     \caption{Channel-wise inlet and outlet pressure distribution (top) and corresponding pressure drop (bottom) for the 3D radial manifold case.}
% \label{fig:pressure_profile_3D_circ}
% \end{figure}

\begin{figure}[htb!]
\centering
\includegraphics[trim=0 0 15 0, clip, width=0.6\linewidth]{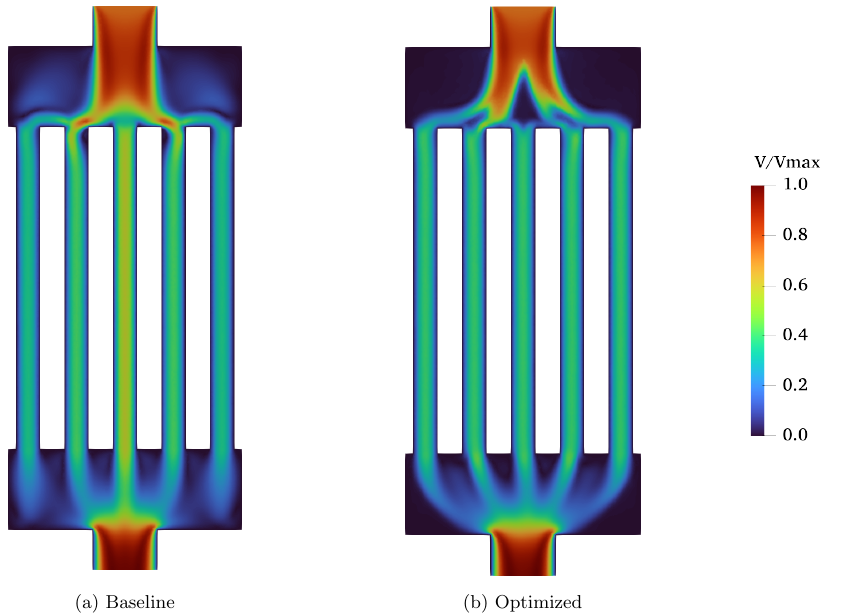}
\caption{Mid-plane, normalized velocity fields for the radial manifold case; (a) Baseline (b) Optimized.}
\label{fig:vel_mid_plane_3D_circ_Case}
\end{figure}

% Velocity Profile
\autoref{fig:vel_mid_plane_3D_circ_Case} presents the mid-plane contours of the normalized velocity magnitude for the baseline and optimized geometries. In the baseline design, peak velocities are concentrated near the central inlet and outlet junctions, while low-velocity and recirculation zones are evident within both manifolds. These features indicate inefficient flow extraction and contribute to elevated local velocity gradients, as well as poor utilization of the outer channels. In contrast, the optimized design exhibits a more uniform velocity distribution across all channels. The flow is more effectively directed at both the inlet and outlet, with reduced stagnation regions and smoother transitions. This leads to improved flow distribution and lower energy dissipation due to more streamlined flow routing.

\section{Conclusions}
The objective of this work was to propose a scalable topology optimization strategy for designing multi-channel flow manifolds with uniform flow distribution. 

To achieve this, a novel design optimization strategy was realized using a density-based topology optimization framework developed within the open-source CFD suite \textit{SU2}. The proposed strategy, referred to as a flow uniformity-based approach (\fua), incorporates the flow maldistribution coefficient as a constraint in the design problem. To evaluate the effectiveness of the proposed strategy, it was benchmarked against two other commonly used design strategies, namely, volume-only approach (\voa) and individual channel approach (\ica). The benchmark study was performed on a planar z-type flow manifold configuration with ten channels, where both the inlet and outlet manifolds were optimized simultaneously. 

Subsequently, to demonstrate the robustness and scalability of the proposed \fua\ strategy, the proposed approach was applied to two exemplary manifold cases in three dimensions. More specifically, a ten channel cylindrical z-type manifold and a radial flow manifold with twenty-five channels; in both cases, the inlet and outlet manifolds were optimized simultaneously.

From the research reported in this manuscript, the following conclusions can be drawn:
\begin{enumerate}
    \item The commonly used \voa\ strategy does not provide an optimum design of the manifold. The \voa\ approach focuses more on reducing the peak velocities in the design domain and hence is suitable to find the least resistance flow path. This is evident from the low pressure drop values for the optimum design obtained using \voa, as compared to \ica\ and \fua\ approaches. Thus, a \voa\ approach cannot be used to design flow manifolds with uniform flow distribution.
    
    \item The optimum designs obtained using the \ica\ and \fua\ approaches lead to identical manifold designs. However, since the \ica\ approach uses constraints for individual channels, the computational cost of one design step scales with the number of channels. Thus making \ica\ computationally expensive, particularly for multi-channel manifold cases. In contrast, the \fua\ approach uses an \review{ensemble quantity}, more precisely the flow maldistribution function, making it independent of the number of channels. Thus making \fua\ approach cost-efficient in realizing the optimum design for multi-channel manifolds.

    \item Achieving uniform flow distribution in a manifold requires a balance with the resulting increase in overall system pressure drop.

    \item Although the \fua\ approach is attractive for designing uniform flow manifolds, it cannot design manifolds that might need custom flow distribution in the channels. In such cases, \ica\ approach can be deemed more efficient as the flow in individual channels can be constrained.
    
    \item Although simple, \fua\ strategy was successful in obtaining feasible optimum manifold designs for the three-dimensional cases featuring ten and twenty-five channels. Therefore by addressing these complex cases, rarely treated in open literature, advocates the robustness, effectiveness and scalability of the proposed \fua\ strategy.
\end{enumerate}

Future research will focus on incorporating turbulence model and heat transfer into the optimization framework, allowing for the simultaneous optimization of flow uniformity and thermal performance. \review{In addition, hybrid formulations that combine the uniformity enforced by FUA with localized ICA on selected outlets could enable overall flow balancing while providing explicit control for critical channels.}
% \clearpage
% % Nomenclature
% \input{Tex/7_Nomenclature}

% \clearpage
% % Credit
% \input{Tex/8_Credit}
% % Declaration
\section*{Declaration of competing interest}
The authors declare that they have no known competing financial interests or personal relationships that could have appeared to influence the work reported in this paper.
% % Acknowledgement
\section*{Acknowledgement}
The research reported in this manuscript was funded by the Vlaanderen Agentschap Innoveren and Ondernemen (VLAIO), Belgium, through the project IAMHEX (grant number HBC.2021.0801). The computational resources and services used in this work were provided by the HPC core facility, CalcUA, of the Universiteit Antwerpen and VSC (Flemish Supercomputer Center), funded by the Research Foundation - Flanders (FWO) and the Flemish Government.
 \bibliographystyle{elsarticle-num} 
 \bibliography{Bib/bibfile}

%% else use the following coding to input the bibitems directly in the
%% TeX file.

% \begin{thebibliography}{00}

% %% \bibitem{label}
% %% Text of bibliographic item

% \bibitem{}

% \end{thebibliography}
\end{document}